\documentclass[aps,prl,superscriptaddress,footinbib,showpacs,twocolumn]{revtex4-2}
\usepackage{comment}

\usepackage{amssymb,latexsym,mathrsfs}
\usepackage{amsfonts}
\usepackage{graphicx}
\usepackage{lmodern}
\usepackage{fix-cm}
\usepackage{xfrac}
\usepackage{color}    
\usepackage{lipsum}
\usepackage{textcomp}
\usepackage{amsmath}
\usepackage{mathtools}
\usepackage{subfigure}
\usepackage{float}
\usepackage{datetime}
\usepackage{hyperref}
 \hypersetup{
           breaklinks=true,   
           colorlinks=true,   
           pdfusetitle=true,  
    linkcolor=red,
    filecolor=magenta,      
    urlcolor=cyan,
        }
\usepackage{tabularx,ragged2e,booktabs}

\newcommand{\be}{\begin{equation}}
\newcommand{\ee}{\end{equation}}
\newcommand{\bearr}{\begin{array}}
\newcommand{\enarr}{\end{array}}

\def\bea{\begin{eqnarray}}
\def\eea{\end{eqnarray}}
\def\ba{\begin{array}}
\def\ea{\end{array}}

\definecolor{dgreen}{rgb}{0,0.7,0}

\begin{document}

\title{Quantum Mpemba effect in a quantum dot with reservoirs}

\author{Amit Kumar Chatterjee}\email{ak.chatterjee@yukawa.kyoto-u.ac.jp}
\affiliation{Yukawa Institute for Theoretical Physics, Kyoto University, Kitashirakawa Oiwakecho, Sakyo-ku, Kyoto 606-8502, Japan}
\author{Satoshi Takada}\email{takada@go.tuat.ac.jp}
\affiliation{Department of Mechanical Systems Engineering and Institute of Engineering, Tokyo University of Agriculture and Technology, 2–24–16 Naka-cho, Koganei, Tokyo 184–8588, Japan}
\author{Hisao Hayakawa}\email{hisao@yukawa.kyoto-u.ac.jp}
\affiliation{Yukawa Institute for Theoretical Physics, Kyoto University, Kitashirakawa Oiwakecho, Sakyo-ku, Kyoto 606-8502, Japan}

\begin{abstract} 
We demonstrate the quantum Mpemba effect in a quantum dot coupled to two reservoirs, described by the Anderson model. We show that the system temperatures starting from two different initial values (hot and cold), cross each other at finite time (and thereby reverse their identities i.e. hot becomes cold and vice versa) to generate {\it thermal} quantum Mpemba effect. The slowest relaxation mode believed to play the dominating role in Mpemba effect in Markovian systems, does not contribute to such anomalous relaxation in the present model. In this connection, our analytical result provides necessary condition for producing quantum Mpemba effect  in the density matrix elements of the quantum dot, as a combined effect of the remaining relaxation modes.  
\end{abstract}

\maketitle

{\it Introduction.-} The Mpemba effect (MPE) is a fascinating counter intuitive phenomenon indicating hot liquid can freeze faster than cold liquid, observed long ago by Aristotle \cite{Aristotle81} and rediscovered by Mpemba and Osborne \cite{Mpemba69}. Various mechanisms have been proposed to explain MPE  \cite{Vynnycky12,Vynnycky15,Mirabedin17,Vynnycky10,Auerbach95,Wojciechowski88,Zhang14,Katz09,Brownridge11,Gijon19}, still lacking any unified theory. In fact, several experiments have raised questions regarding the validity of MPE \cite{Burridge16}. A problem regarding the correct definition of MPE for various systems \cite{Mpemba69,Chaddah10,Ahn16,Hu18}   is the complexity of the phase transitions associated with these cases. In this connection, the experimental observation of MPE in colloidal system without phase transition is remarkable \cite{Kumar20}.    

Although MPE has been originally perceived as a thermal phenomena of anomalous cooling in liquids, later it is identified as more general anomalous relaxation occurring in wide variety of systems, including colloids \cite{Kumar20,Kumar22}, granular gases \cite{Santos17,Torrente19,Biswas20,Biswas21,Mompo20,Megias22b}, optical resonators \cite{Keller18,Santos20,Patron21}, inertial suspensions \cite{Takada21a,Takada21b}, Markovian models \cite{Lu17, Klich19, Busiello21,Lin22} and others \cite{Greaney11,Baity-Jesi19,Gonzalez2021,Yang20,Yang22,Gonzalez21,Chetrite21,Holtzman22}. To analyze MPE in classical systems, time variations of temperature \cite{Takada21a,Biswas21},  energy \cite{Biswas20}, viscocity \cite{Takada21a} have been employed in granular gases and inertial suspensions whereas entropic distance-from-equilibrium functions have been applied to Markov jump processes \cite{Lu17,Klich19} and colloidal systems \cite{Kumar20}. Recently, nontrivial connections between  thermal and entropic MPEs have been exploited \cite{Megias22a}  and the crucial dependence of MPE on the choice of the observables  is studied \cite{Biswas23}.  

In spite of substantial works on the classical MPE, investigations on quantum Mpemba effect (QMPE) have been few \cite{Nava19,Carollo21,Manikandan21,Ares23,Ivander23}. The studies of QMPE have been based on the entropic distance-from-equilibrium functions   \cite{Carollo21,Manikandan21,Ivander23},  entanglement asymmetry \cite{Ares23} and  magnetization \cite{Nava19}. Importantly, QMPE lacks the  analysis of temperature and therefore the notion of thermal QMPE is missing. Secondly, the criterion for QMPE \cite{Carollo21,Ivander23} and  MPE \cite{Lu17,Klich19,Chetrite21} in both quantum and classical  Markovian systems solely focus on the slowest relaxation mode. Complete absence of the slowest relaxation mode for certain parameter choices or initial conditions leads to exponentially faster relaxation, called {\it strong} MPE \cite{Klich19,Walker21,Kumar20} and {\it strong} QMPE \cite{Carollo21}. However, the roles of other relaxation modes in generating Markovian MPE and QMPE remain unexplored. 

In this Letter, we address the above mentioned issues: (i) possibility of thermal QMPE, and (ii) the role of relaxation modes other than the slowest one in producing QMPE in density matrix elements and temperature. For the demonstration of QMPE, we examine a quantum dot system coupled to two reservoirs, described by the Anderson model. The occurrence of QMPE is defined as the {\it finite time crossing} of temporal trajectories of any entity starting from two different initial conditions that reach the same steady state. We show that the system temperature exhibits thermal QMPE with the variation of control parameters. Moreover, the slowest relaxation mode does not contribute to the QMPE in the present model and  we illustrate the combined role of the remaining eigenmodes in generating QMPE. 

{\it Model.-} We consider a single level quantum dot (QD) coupled to two reservoirs ($\mathrm{L}$ and $\mathrm{R}$). The total system is described by the Anderson model with the Hamiltonian $\hat{H}_{\mathrm{tot}}=\hat{H}_{\mathrm{s}}+\hat{H}_{\mathrm{r}}+\hat{H}_{\mathrm{int}}$ where $\hat{H}_{\mathrm{s}},\hat{H}_{\mathrm{r}}$ are the Hamiltonians for the QD and the two reservoirs respectively and $\hat{H}_{\mathrm{int}}$ denotes the system-reservoirs interaction. The explicit forms of the Hamiltonians are:  \cite{Yoshii13,Hayakawa21},
\begin{eqnarray}
\hat{H}_{\mathrm{s}}=\sum_{\sigma}\epsilon_0 \hat{d}^\dagger_\sigma\hat{d}_\sigma\,&+&\,U \hat{n}_\uparrow\hat{n}_\downarrow,\hspace*{0.1 cm} 
 \hat{H}_{\mathrm{r}}=\sum_{\gamma,k,\sigma} \epsilon_k \hat{a}^\dagger_{\gamma,k,\sigma}\hat{a}_{\gamma,k,\sigma}\cr
\hat{H}_{\mathrm{int}}&=& \sum_{\gamma,k,\sigma} V_\gamma \hat{d}^\dagger_{\sigma}\hat{a}_{\gamma,k,\sigma}\,+\, \mathrm{h.c.}
\label{eq:and2}
\label{eq:Hamiltonian} 
\end{eqnarray}
The parameters $\epsilon_0$ and $\epsilon_k$ correspond to an electron energy in the QD and the reservoirs respectively, $U$ is the electron-electron repulsion energy in the QD. The index $\sigma$ denotes up-spin ($\uparrow$) and down-spin ($\downarrow$), $\gamma$ represents $\mathrm{L}$ and $\mathrm{R}$. The creation (annihilation) operators for the QD and the reservoirs are $\hat{d}^\dagger$ ($\hat{d}$) and $\hat{a}^\dagger$ ($\hat{a}$), respectively. Here $\hat{n}_{\sigma}=\hat{d}_{\sigma}^\dagger\hat{d}_{\sigma}$ is the number operator. The coupling strength between QD and $\mathrm{L}$ ($\mathrm{R}$) is $V_{\mathrm{L}}$ ($V_{\mathrm{R}}$). We adopt a model in the {\it wide band limit} for reservoirs \cite{Yoshii13,Hayakawa21,Yoshii22,Nakajima15}. We denote the line width $\Gamma=\pi\Omega V^2$ $(V^2 = V_{\mathrm{L}}^2 + V_{\mathrm{R}}^2)$ where $\Omega$ is the density of states in the reservoirs. The chemical potentials of $\mathrm{L}$ $(\mathrm{R})$ is $\mu_\mathrm{L}$ $(\mu_\mathrm{R})$ and their temperatures are taken equal $T_\mathrm{L}=T_\mathrm{R}=T$. We consider temperatures much higher than the Kondo temperature to ignore the Kondo effect  \cite{Cronenwett98,Sakano06,Hiroka17}. We focus on the weak coupling to disregard co-tunneling \cite{Keller16}.

The tunneling between QD and the reservoirs means the QD has four possible states, namely the doubly occupied state ($ \uparrow\downarrow$), singly occupied up-spin state ($\uparrow$), singly occupied down-spin state ($\downarrow$) and empty state;  enumerated by $\alpha=1,2,3,4$, respectively. Correspondingly, the elements of the density matrix operator $\hat{\rho}$  for the QD are represented by $\rho_\alpha$. Using wide band approximation, $\hat{\rho}$ reduces to a purely diagonal form \cite{Yoshii13,Hayakawa21,Yoshii22,Nakajima15} with four non-zero elements $\rho_\alpha$. The dynamics of the QD is described by the quantum Master equation 
\begin{equation}
\frac{d}{d\tau}\hat{\rho}=\hat{K}\hat{\rho},
\label{eq:qme} 
\end{equation}
 where $\tau:=\Gamma t$ is the dimensionless time. The transition matrix $\hat{K}$ has the form \cite{Yoshii13,Nakajima15}
\begin{equation}
\hat{K}:=\left(\begin{array}{cccc}
               -2f_-^{(1)} & f_+^{(1)} & f_+^{(1)} & 0 \\
               f_-^{(1)} & -f_-^{(0)}-f_+^{(1)} & 0 & f_+^{(0)} \\
               f_-^{(1)} & 0 & -f_-^{(0)}-f_+^{(1)} & f_+^{(0)} \\
               0 & f_-^{(0)} & f_-^{(0)} & -2f_+^{(0)} \\
              \end{array}
\right).
\label{eq:transition_matrix} 
\end{equation}
The factors $f_{\pm}^{(j)}$ ($j=0,1$)  [Eq.~(\ref{eq:transition_matrix})] are related to the physical input parameters as  
\begin{eqnarray}
&& f_+^{(j)}:=f_{\mathrm{L}}^{(j)}(\mu_\mathrm{L},U,\epsilon_0,T)+f_{\mathrm{R}}^{(j)}(\mu_\mathrm{R},U,\epsilon_0,T), \hspace*{0.2 cm} j=0,1\cr
&& f_\gamma^{(j)}(\mu_\gamma,U,\epsilon_0,T):=\frac{1}{1+e^{(\epsilon_0+jU-\mu_\gamma)/T}} \hspace*{0.5 cm} \gamma=\mathrm{L},\mathrm{R},
\label{eq:fermi_factors} 
\end{eqnarray}
with $f_{\mathrm{L},\mathrm{R}}^{(j)}$ ($j=0,1$) being the Fermi-Dirac distribution. Among the parameters $f_{\pm}^{(j)}$ ($j=0,1$) in Eq.~(\ref{eq:transition_matrix}), only two are independent (we consider $f_+^{(j)}$ with $j=0,1$) because $f_-^{(j)}=2-f_+^{(j)}$. The exact analysis of QMPE in the QD can be performed in terms of $f_+^{(0)}$ and $f_+^{(1)}$ and  Eq.~(\ref{eq:fermi_factors}) is used to express the results in terms of chemical potential, temperature and the Hamiltonian parameters.

{\it Protocol.-} We use two different sets $\mathrm{I}$ and $\mathrm{II}$ of initial conditions,  $\hat{\rho}^\mathrm{I}(\tau=0)$ and 
$\hat{\rho}^\mathrm{II}(\tau=0)$. Both these initial conditions are chosen in the form of the steady state distribution corresponding to the largest (zero) eigenvalue of $\hat{K}$ [Eq.~(\ref{eq:transition_matrix})]. They may differ in the values of one or more input parameters. For initial condition $\mathrm{I}$, we choose $\mu_\mathrm{L}^\mathrm{I}\neq\mu_\mathrm{R}^\mathrm{I}$, whereas for $\mathrm{II}$, we take $\mu_\mathrm{L}^\mathrm{II}=\mu_\mathrm{R}^\mathrm{II}=\mu^{\mathrm{II}}$. For both initial conditions, the reservoirs are maintained at the same initial temperature $T_\mathrm{i}$. At $\tau=0$, we perform {\it instantaneous quench} for both initial conditions such that their chemical potentials are quenched to $\mu$ and  $T_\mathrm{i}$ is quenched to $T$. The parameters after quench can be higher or lower than their values before quench. We follow the time evolution of an entity for $\mathrm{I}$ and $\mathrm{II}$, to see if they cross each other at some finite time $\tau>0$ before reaching the same steady state, and thereby if that entity  exhibits QMPE. We investigate QMPE in the density matrix elements, and in the temperature to explicitly survey the possibility of thermal QMPE.

{\it QMPE in density matrix elements.-} The time evolution of $\rho_\alpha(\tau)$ ($\alpha=1,2,3,4$) is governed by the eigenvalues and (right and left) eigenvectors of $\hat{K}$ [Eq.~(\ref{eq:transition_matrix})]. The formal expressions of $\rho_\alpha(\tau)$, for $\mathrm{I}$ and $\mathrm{II}$, are
\begin{eqnarray}
\rho^{\mathrm{I}}_\alpha(\tau)\,&=&\, \sum_{n=1}^{4} e^{\lambda_n \tau} \hat{R}_{\alpha,n}a_{n}^{\mathrm{I}}, \hspace*{0.2 cm} a_{n}^{\mathrm{I}}=\,\sum_{m=1}^{4}\widehat{L}_{n,m}\rho^{\mathrm{I}}_{m}(0), \cr
\rho^{\mathrm{II}}_\alpha(\tau)\,&=&\, \sum_{n=1}^{4} e^{\lambda_n \tau} \hat{R}_{\alpha,n}a_{n}^{\mathrm{II}}, \hspace*{0.2 cm} a_{n}^{\mathrm{II}}=\,\sum_{m=1}^{4}\widehat{L}_{n,m}\rho^{\mathrm{II}}_{m}(0),
\label{eq:rho_alpha}
\end{eqnarray}
where $\lambda_n$ ($n=1,2,3,4$) are the eigenvalues of $\hat{K}$ such that $\lambda_1>\lambda_2>\lambda_3>\lambda_4$. Here $\lambda_1$ ($=0$) and $\lambda_2$ correspond to the steady state and the slowest relaxation mode, respectively. The matrices $\hat{R}$ and $\hat{L}$ consist of the right eigenvectors and left eigenvectors of $\hat{K}$, respectively \cite{supp}.  The coefficients $a^{\mathrm{I,II}}_n$ [Eq.~(\ref{eq:rho_alpha})] contain the effects of the initial conditions. In many literature, only the coefficient $a^{\mathrm{I,II}}_2$, corresponding to the slowest eigenmode, is considered to analyze the MPE at sufficiently large time where the effects of other coefficients $a^{\mathrm{I,II}}_n$ $(n>2)$ are assumed negligible \cite{Lu17,Chetrite21,Carollo21,Ivander23}.  Several studies concentrate on engineering special initial conditions that lead to $a_2=0$ \cite{Carollo21,Walker21} causing strong MPE with exponentially faster relaxation. In this connection, we have found an intriguing fact that the Anderson model of QD satisfies the condition 
\begin{equation}
a_{2}^{\mathrm{I}}=a_{2}^{\mathrm{II}}=0,
\label{eq:a2_0} 
\end{equation}
irrespective of the particulars of initial conditions and  parameter values. The slowest relaxation mode $\lambda_2$ does not contribute to the time evolved density matrix elements and other observables, for any initial condition distributed in the steady state form. Thus, the coefficient $a^{\mathrm{I,II}}_2$ is not appropriate to discuss QMPE for the Anderson model. 

This naturally raises the question: what are the roles of the coefficients $a_3$ and $a_4$, corresponding to the remaining relaxation modes $n=3,4$, in producing QMPE ? To answer this, we consider the difference between the time evolutions of density matrix elements starting from $\mathrm{I}$ and $\mathrm{II}$, i.e.
\begin{equation}
\Delta\rho_\alpha(\tau):=\rho^{\mathrm{I}}_\alpha(\tau)-\rho^{\mathrm{II}}_\alpha(\tau), \hspace*{0.5 cm} \alpha=1,2,3,4.
\label{eq:delta_rho_1}
\end{equation}
To obtain QMPE, we have to make sure that some $\Delta\rho_\alpha(\tau_\alpha)=0$ at finite time $\tau=\tau_\alpha$. We obtain the following analytical expressions for $\Delta\rho_\alpha(\tau)$:
\begin{equation}
\Delta\rho_\alpha(\tau)=e^{\lambda_3 \tau}\hat{R}_{\alpha,4}\Delta a_4\left[S_\alpha\,+\,e^{-(\lambda_3-\lambda_4)\tau}\right],
\label{eq:delta_rho_2} 
\end{equation}
where $S_\alpha:=(\hat{R}_{\alpha,3}\Delta a_3)/(\hat{R}_{\alpha,4}\Delta a_4)$ and $\Delta a_\alpha:=a^{\mathrm{I}}_\alpha-a^{\mathrm{II}}_\alpha$. Since, the expression of $S_\alpha$ explicitly depends on the ratio $\Delta a_3/\Delta a_4$, we conclude that QMPE in the density matrix elements is dictated by {\it both} the surviving relaxation modes rather than only one of them. Since in Eq.~(\ref{eq:delta_rho_2}), $0\leqslant e^{-(\lambda_3-\lambda_4)\tau}\leqslant1$, the {\it necessary} condition  to ensure QMPE in $\rho_\alpha(t)$ is 
\begin{eqnarray}
&& S_\alpha<0 \hspace*{0.3 cm} \& \hspace*{0.3 cm} |S_\alpha|<1.
\label{eq:criterion} 
\end{eqnarray}
Note that we have not used the explicit expressions of the eigenvalues and eigenvectors of $\hat{K}$ to derive the criterion Eq.~(\ref{eq:criterion}). 
One can control one or more parameters from $(\mu_{\mathrm{L}}^{\mathrm{I}},\mu_{\mathrm{R}}^{\mathrm{I}},\mu^{\mathrm{II}},T_{\mathrm{i}};\mu,T)$. We choose to vary $\mu_{\mathrm{L}}^{\mathrm{I}}$ or $\mu_{\mathrm{R}}^{\mathrm{I}}$ or both. We denote the number of density matrix elements showing QMPE by $\nu(\hat{\rho})$ which can take one of the four possible values $0,1,2$ and $3$. 
\begin{figure}[t]
  \centering \includegraphics[width=8.6 cm]{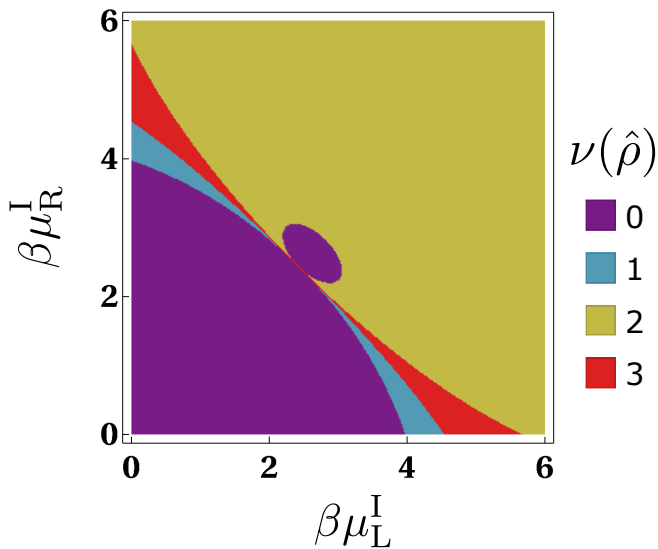}
\caption{The figure illustrates different regions in the $\beta\mu_{\mathrm{L}}^{\mathrm{I}}-\beta\mu_{\mathrm{R}}^{\mathrm{I}}$ plane with distinct values of $\nu(\hat{\rho})$ (number of density matrix elements showing QMPE). Parameters used are $\beta\epsilon_0=2.0, \beta U=1.25, \beta\mu^{\mathrm{II}}=2.43, \beta T_\mathrm{i}=1.15, \beta\mu=2.0$.}
\label{fig:nu_rho}
\end{figure}

In Fig.~\ref{fig:nu_rho}, we present the variation of $\nu(\hat{\rho})$ in the $\beta\mu_{\mathrm{L}}^{\mathrm{I}}-\beta\mu_{\mathrm{R}}^{\mathrm{I}}$ plane, where $\beta=1/T$. This figure is constructed by directly implementing the criterion in Eq.~(\ref{eq:criterion}). The  parameter plane captures all possible values of $\nu(\hat{\rho})$.  
The behavior of $\nu(\hat{\rho})$ is naturally symmetric with respect to $\beta\mu_{\mathrm{L}}^{\mathrm{I}}$ and $\beta\mu_{\mathrm{R}}^{\mathrm{I}}$. A quadrant centering $\beta\mu_{\mathrm{L}}^{\mathrm{I}}=\beta\mu_{\mathrm{R}}^{\mathrm{I}}=0$ appears on the plane that forbids the occurrence of QMPE and is characterized by $\nu(\hat{\rho})=0$. As we move away from this quadrant, the density matrix elements starts showing QMPE. If we fix one of the parameters $\beta\mu_{\mathrm{L}}^{\mathrm{I}}$ and $\beta\mu_{\mathrm{R}}^{\mathrm{I}}$ and increase the other, the value of $\nu(\hat{\rho})$ does not  increase monotonically in the order of $\nu(\hat{\rho})=0,1,2,3$, rather we have a narrow parameter region exhibiting $\nu(\hat{\rho})=3$ sandwiched between regions showing $\nu(\hat{\rho})=1$ and $\nu(\hat{\rho})=2$.
The fact that the parameter regions showing  $\nu(\hat{\rho})=3$ or $1$ are much narrower than the regions displaying $\nu(\hat{\rho})=2$ or $0$, is generic for our model \cite{supp}. 
\begin{figure}[t]
  \centering \includegraphics[width=8.6 cm]{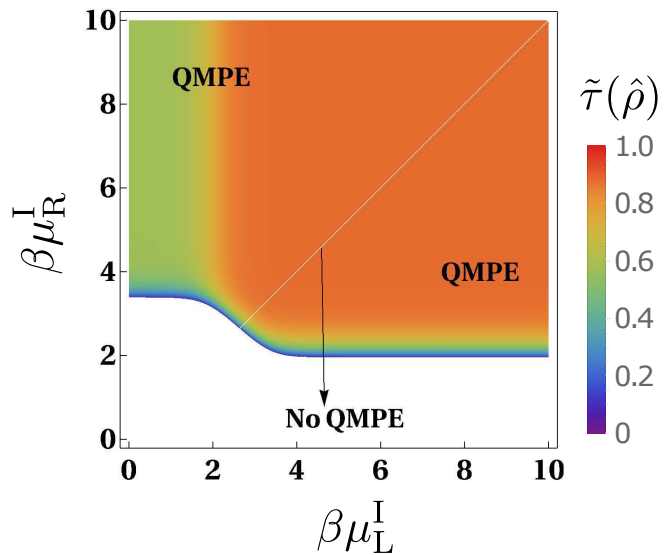}
  \caption{The figure shows the QMPE for $\hat{\rho}$ in $\beta\mu_{\mathrm{L}}^{\mathrm{I}}-\beta\mu_{\mathrm{R}}^{\mathrm{I}}$ plane characterized by $\tilde{\tau}(\hat{\rho})$ [Eq.~(\ref{eq:order_parameter})]. The white regions represent the absence of QMPE. Parameters used are $\beta\epsilon_0=2.0, \beta U=1.25, \beta T_\mathrm{i}=0.25, \beta\mu=2.0$ and $\beta\mu^{\mathrm{II}}=\beta\mu_{\mathrm{R}}^{\mathrm{I}}$.}
\label{fig:tau_rho}
\end{figure}

Since QMPE is a dynamical phenomenon, we characterize the occurrence of QMPE in $\hat{\rho}$ by the {\it temporal} order parameter defined below, 
\begin{eqnarray}
\tilde{\tau}(\hat{\rho})&=& \mathrm{max}[\tau_1,\tau_2,\tau_3,\tau_4]\,\,\, \mathrm{if} \,\,\, 0<\tau_\alpha<\infty, \cr
\tilde{\tau}(\hat{\rho})&\to& \infty \,\,\, \mathrm{if} \,\,\, \mathrm{no}\,\, \mathrm{finite}\,\, \tau_\alpha\,\,\mathrm{exists}\,\,\,\mathrm{\forall \alpha}, 
\label{eq:order_parameter} 
\end{eqnarray}
where $\tau_\alpha$ ($\alpha=1,2,3,4$) is the solution of 
$\Delta \rho_\alpha(\tau_\alpha)=0$, and $\mathrm{max}[x_1,x_2,x_3,x_4]$ selects the largest $x_i$ among $x_1,x_2,x_3$ and $x_4$. The reason to focus on the maximum among $\tau_\alpha$-s in Eq.~(\ref{eq:order_parameter}) is to detect the largest time  which bears the memory effect from the initial quench. The trivial steady state solution $\Delta \rho_\alpha(\tau\rightarrow\infty)=0$ $\forall \alpha$ must be avoided. In Fig.~\ref{fig:tau_rho}, we present the behavior of $\tilde{\tau}(\hat{\rho})$ in the $\beta\mu_{\mathrm{L}}^{\mathrm{I}}-\beta\mu_{\mathrm{R}}^{\mathrm{I}}$ plane. The  majority of the parameter region exhibits QMPE with largest $\tilde{\tau}(\hat{\rho})$ of the order of unity. The proposed order parameter provides prominent boundaries demarcating regions with and without QMPE. To reduce the number of independent parameters we consider $\beta\mu^{\mathrm{II}}=\beta\mu_{\mathrm{R}}^{\mathrm{I}}$ in Fig.~\ref{fig:tau_rho}. However, such special scenario creates asymmetric nature of $\tilde{\tau}(\hat{\rho})$ with respect to $\beta\mu_{\mathrm{L}}^{\mathrm{I}}$ and $\beta\mu_{\mathrm{R}}^{\mathrm{I}}$, evident in Fig.~\ref{fig:tau_rho}. Notably, for fixed $\beta\mu_{\mathrm{R}}^{\mathrm{I}}<2$, QMPE is prohibited in the whole tunable range of  $\beta\mu_{\mathrm{L}}^{\mathrm{I}}$ ($0$ to $\infty$).

{\it QMPE in system temperature.-} Next we investigate the possibility of QMPE in the system temperature, i.e., quantum counterpart of the original MPE. The concept of temperature has to be {\it dynamical} as we study the temporal relaxation of the system. The process of thermalization itself can be tricky for quantum systems \cite{Nandkishore15,Mori18,Shiraishi21}. Nevertheless, we use the definition of time dependent temperature $T_{\mathrm{s}}(\tau)$ \cite{Ali20} as
\begin{equation}
T_{\mathrm{s}}(\tau):=\frac{\partial E_{\mathrm{s}}(\tau)}{\partial S_{\mathrm{vN}}(\tau)}=\left.\frac{\partial  E_{\mathrm{s}}(\tau)}{\partial \tau}\right/\frac{\partial S_{\mathrm{vN}}(\tau)}{\partial \tau},
\label{eq:temperature} 
\end{equation}
where  $S_{\mathrm{vN}}(\tau)=-\sum_{\alpha} \rho_\alpha(\tau)\mathrm{ln}(\rho_\alpha(\tau))$ is the von-Neumann entropy and $E_{\mathrm{s}}(\tau)=\mathrm{Tr}[\hat{\rho}(\tau) \hat{H}_\mathrm{s}]$ is the average energy. The thermalization and the validity of the system temperature are  ensured from the checked fact that different initial conditions converge to the identical steady state value $T_{\mathrm{s}}(\tau\rightarrow\infty)$. Also, the classical limit is recovered as $T_{\mathrm{s}}(\tau\rightarrow\infty)$ matches with the reservoir temperature $T$ when the Fermi-Dirac distribution [Eq.~(\ref{eq:fermi_factors})] can be approximated as Maxwell-Boltzmann distribution \cite{supp}.
\begin{figure}[t]
\centering \includegraphics[width=8.6 cm]{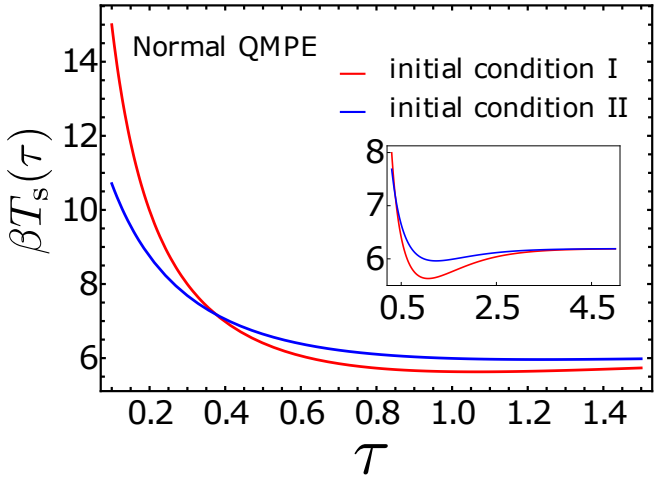}
\caption{The figure shows thermal QMPE where temperatures [Eq.~(\ref{eq:temperature})] starting from two different initial conditions cross each other at finite time. The inset shows their convergence to the same steady state. Parameters used are $\beta\epsilon_0=2.0, \beta U=1.25, \beta\mu_{\mathrm{L}}^{\mathrm{I}}=4.5, \beta\mu_{\mathrm{R}}^{\mathrm{I}}=1.0, \beta\mu^{\mathrm{II}}=2.43, \beta T_\mathrm{i}=1.15, \beta\mu=2.0$.}
\label{fig:temperature}
\end{figure}

Interestingly, in Fig.~\ref{fig:temperature}, we observe that $\beta T_{\mathrm{s}}(\tau)$  starting from two different initial values cross each other at a finite time showing QMPE. Since both the  initial temperatures are higher than the steady state value, this QMPE involves cooling processes where the initially hotter system becomes colder after the crossing, and thereby produces the {\it normal} QMPE \cite{Takada21a}. It is fascinating that the QD indeed generates thermal MPE. The inset of Fig.~\ref{fig:temperature} confirms that both initial temperatures reach same steady state.
\begin{figure}[t]
  \centering \includegraphics[width=8.6 cm]{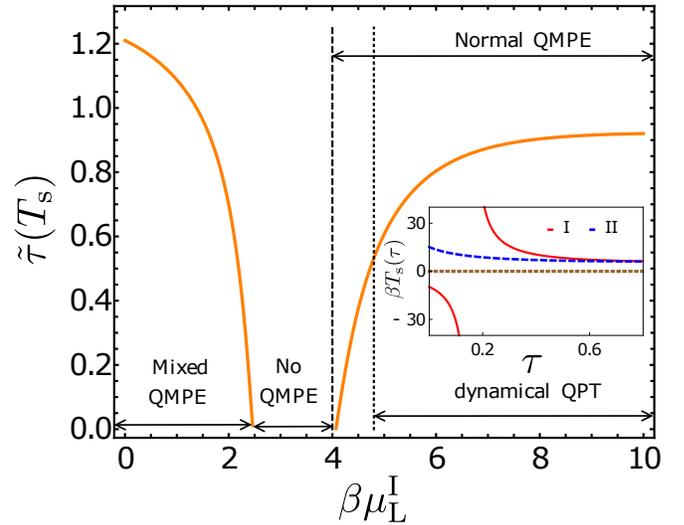}
  \caption{The figure shows variation of $\tilde{\tau}(T_{\mathrm{s}})$ [Eq.~(\ref{eq:tau_temperature})] characterizing thermal QMPE. We observe mixed QMPE and normal QMPE, separated by a region with no thermal QMPE. A majority of the parameter range showing normal QMPE is associated with dynamical QPT (inset). Parameters used are $\beta\epsilon_0=2.0, \beta U=1.25, \beta\mu_{\mathrm{R}}^{\mathrm{I}}=1.0, \beta\mu^{\mathrm{II}}=2.43, \beta T_\mathrm{i}=1.15, \beta\mu=2.0$.}
\label{fig:tau_temperature}
\end{figure}

We should examine if the occurrence of thermal QMPE in Fig.~\ref{fig:temperature} is an isolated incident in a rather large parameter space. For this purpose, analogous to $\tilde{\tau}(\hat{\rho})$ [Eq.~(\ref{eq:order_parameter})], we define $\tilde{\tau}(T_{\mathrm{s}})$ characterizing thermal QMPE as 
\begin{eqnarray}
0<\tilde{\tau}(T_{\mathrm{s}})<\infty&:& \hspace*{0.2 cm} \mathrm{thermal}\,\,\mathrm{QMPE},\cr
\tilde{\tau}(T_{\mathrm{s}})\to\infty&:& \hspace*{0.2 cm} \mathrm{no}\,\,\mathrm{QMPE},
\label{eq:tau_temperature} 
\end{eqnarray}
where $\tilde{\tau}(T_{\mathrm{s}})$ is the solution of $\Delta T_\mathrm{s}(\tilde{\tau})=0$ with $\Delta T_\mathrm{s}:=T_{\mathrm{s}}^{\mathrm{I}}-T_{\mathrm{s}}^{\mathrm{II}}$. In Fig.~\ref{fig:tau_temperature}, we present the behavior of $\tilde{\tau}(T_{\mathrm{s}})$ with the variation of $\beta\mu_{\mathrm{L}}^{\mathrm{I}}$. This remarkably rich diagram reveals that thermal QMPE is rather generic than occasional. We observe large parameter ranges exhibiting normal QMPE (an example being Fig.~\ref{fig:temperature}) and {\it mixed} QMPE (one of the initial temperatures is lower than the steady state value) \cite{Takada21a}, with intermediate region displaying absence of QMPE. The model is also capable to exhibit {\it inverse} QMPE \cite{supp} (both initial temperatures are lower than the steady state value \cite{Santos17}), although the inverse QMPE is much weaker than normal and mixed QMPEs. An interesting observation is that most of the parameter region demonstrating normal QMPE are associated with non-analyticity in the system temperature  (Fig.~\ref{fig:tau_temperature} inset) resulting in dynamical quantum phase transition (QPT) \cite{Heyl18}. The negative temperature (Fig.~\ref{fig:tau_temperature} inset) originates from the {\it non-monotonicity} of the entropy while the energy changes monotonically. Negative temperatures along with bounded energy spectrum have been predicted \cite{Onsager49,Landau80} and experimentally observed \cite{Gauthier19,Johnstone19} for two-dimensional vortices,  localized spin systems \cite{Ramsey56,Purcell51,Medley11}, bosonic single mode cavity \cite{Ali20} etc. The thermal QMPE in the present model remains unaffected by the dynamical QPT as $\tilde{\tau}(T_{\mathrm{s}})$ increases continuously across the borderline demarcating the presence and absence of dynamical QPT.

Apart from temperature and density matrix, energy, von-Neumann entropy and Kullback-Leibler divergence ($D_{\mathrm{KL}}(\tau)$) can also exhibit QMPE \cite{supp}. Remarkably, we have found parameter regions where thermal QMPE is observed but $D_{\mathrm{KL}}(\tau)$ does not show QMPE and vice versa, implying that here $D_{\mathrm{KL}}(\tau)$ cannot act as an alternative indicator for thermal QMPE \cite{supp}. We observe that higher initial difference between initial and steady state, goes faster to zero for QMPE in energy and entropy \cite{supp}. Similar observation regarding faster restoration of more initially broken symmetry has been studied recently using entanglement asymmetry \cite{Ares23}.

{\it Summary.-} We have demonstrated QMPE in a  single level quantum dot coupled to two reservoirs, described by the Anderson model. Interestingly, the slowest relaxation mode which has been by far the only focus for producing MPE in Markovian systems, has no contribution to QMPE for the Anderson model. Rather, we have presented the necessary criterion, involving the combination of remaining relaxation modes, to obtain QMPE in the density matrix elements. We have achieved the thermal QMPE in the temperature where an initially hotter system can cool faster than an initially colder system and they reverse their identities (hotter becomes colder and vice versa) after some finite time. 

It would be important to investigate the thermal QMPE in other  quantum systems and systematically establish the general framework for it. We wish to explore the connection of symmetry breaking to QMPE \cite{Ares23} in quantum dots. There have been several experiments on QD \cite{Winkler05,Feve07,Blumenthal07,Connolly13,Liu21,Zanten16,Corral20}, including single-level QD coupled to two leads with different chemical potentials and controlled by gate voltages \cite{Zanten16,Corral20}. Thus, it is straightforward to perform experiments on QMPE by controlling chemical potentials discussed in this Letter. 

{\it Acknowledgements.-} We thank Fr\'{e}d\'{e}ric van Wijland, Raphael Chetrite, Marija Vucelja, Ryo Hanai, Manas Kulkarni, Amit Dey and Brett Min for useful discussions. This work is partially supported by the Grants-in-Aid for Scientific Research (Grant No. 21H01006 and No. 20K14428). A.K.C. gratefully acknowledges postdoctoral fellowship from the YITP.    The numerical calculations have been done on Yukawa-21 at the YITP.

\newpage
\setcounter{equation}{0}
\setcounter{figure}{0}

\renewcommand{\theequation}{S\arabic{equation}}
\renewcommand{\thefigure}{S\arabic{figure}}
\renewcommand{\bibnumfmt}[1]{[S#1]}
\renewcommand{\citenumfont}[1]{S#1}

\onecolumngrid
\clearpage

\section*{Supplemental Material}
The supplemental material explains the details of the calculations and additional physical features regarding QMPE in the QD, that are not included in the main text. In Sec.~\ref{transition-matrix} we provide the explicit expressions for the eigenvalues and eigenvectors of the transition matrix $\hat{K}$ that are used to calculate the time evolution of the density matrix elements. In Sec.~\ref{phase-diagram}, we show that the analysis of QMPE in density matrix elements in the multi-dimensional physical parameter space can be simplified dramatically by using two variables only, $\eta_{\mathrm{bq}}$ [combination of control parameters only before quench (bq)] and $\eta_{\mathrm{aq}}$ [combination of control parameters only after quench (aq)]. We classify different regions in the whole $\eta_{\mathrm{aq}}-\eta_{\mathrm{bq}}$ plane characterized by $\nu(\hat{\rho})=0,1,2,3$. In Sec.~\ref{mixed-inverse}, we present examples of thermal QMPE which are either {\it mixed} or {\it inverse} in nature. The validity of the system  temperature used to analyze thermal QMPE is discussed in context of recovery of the classical limit in Sec.~\ref{classical}. We provide examples of QMPE in energy, von-Neumann entropy and Kullback-Leibler divergence in Sec.~\ref{energy-entropy}.

\section{Eigenvalues and eigenvectors of the transition rate matrix $\hat{K}$}
\label{transition-matrix}
We start from the quantum Master equation given by Eqs.~(2) and (3) of the main text,  describing the time evolution of the density matrix for the quantum dot. The four eigenvalues of $\hat{K}$ are given by
\begin{equation}
\lambda_1=0, \hspace*{ 1 cm} \lambda_2=-2+f_+^{(0)}-f_+^{(1)},\hspace*{ 1 cm} \lambda_3=-2-f_+^{(0)}+f_+^{(1)},\hspace*{ 1 cm}\lambda_4=-4.
\label{eq:and9} 
\end{equation}
Keeping in mind that $f_+^{(0)}>f_+^{(1)}$ for all possible choices of chemical potentials and temperatures, we understand that
\begin{eqnarray}
\mathrm{steady}\,\,\mathrm{state}\,\,:\,\,&& \lambda_1=0 \cr
\mathrm{slowest}\,\,\mathrm{relaxation}\,\,\mathrm{mode}:\,\, &&\lambda_2=-2+f_+^{(0)}-f_+^{(1)}.
\label{eq:and10} 
\end{eqnarray}
The steady state $|\hat{\rho}_{\lambda_1}\rangle=|\hat{\rho}^{ss}(f_+^{(0)},f_+^{(1)})\rangle$ is given by the right eigenvector (normalized) corresponding to the eigenvalue zero
\begin{equation}
|\hat{\rho}^{ss}(f_+^{(0)},f_+^{(1)})\rangle =\frac{1}{4+2(f_+^{(0)}-f_+^{(1)})}\left(\begin{array}{c}
f_+^{(0)}f_+^{(1)} \\
f_+^{(0)}(2-f_+^{(1)}) \\
f_+^{(0)}(2-f_+^{(1)}) \\
(2-f_+^{(0)})(2-f_+^{(1)}) \\
\end{array}
\right). \label{eq:and11}
\end{equation}
Note that $\rho^{ss}_{\uparrow}=\rho^{ss}_{\downarrow}$, i.e. the singly occupied up-spin and down-spin state occur with same probability, there is no external field to differentiate between these two states. 

The matrix $\hat{R}$ containing the (normalized) right eigenvectors of the Lindbladian as columns, is given by
\begin{equation}
\hat{R}=\left(\begin{array}{cccc}
\frac{f_+^{(0)}f_+^{(1)}}{4+2(f_+^{(0)}-f_+^{(1)})} & 0 & \frac{2f_+^{(0)}f_+^{(1)}}{-4+\left(f_+^{(0)}-f_+^{(1)}\right)^2} & \frac{f_+^{(0)}f_+^{(1)}}{4-2(f_+^{(0)}-f_+^{(1)})}\\
\frac{f_+^{(0)}(2-f_+^{(1)})}{4+2(f_+^{(0)}-f_+^{(1)})} & -\frac{1}{2}  & -\frac{f_+^{(0)}(2-f_+^{(0)}-f_+^{(1)})}{-4+\left(f_+^{(0)}-f_+^{(1)}\right)^2} & -\frac{f_+^{(0)}f_+^{(1)}}{4-2(f_+^{(0)}-f_+^{(1)})}\\
\frac{f_+^{(0)}(2-f_+^{(1)})}{4+2(f_+^{(0)}-f_+^{(1)})} & \frac{1}{2}   & -\frac{f_+^{(0)}(2-f_+^{(0)}-f_+^{(1)})}{-4+\left(f_+^{(0)}-f_+^{(1)}\right)^2} & -\frac{f_+^{(0)}f_+^{(1)}}{4-2(f_+^{(0)}-f_+^{(1)})}\\
\frac{(2-f_+^{(0)})(2-f_+^{(1)})}{4+2(f_+^{(0)}-f_+^{(1)})} & 0  & -\frac{2f_+^{(0)}(2-f_+^{(0)})}{-4+\left(f_+^{(0)}-f_+^{(1)}\right)^2} & \frac{f_+^{(0)}f_+^{(1)}}{4-2(f_+^{(0)}-f_+^{(1)})}\\
\end{array}
\right).
\label{eq:and12} 
\end{equation}
The matrix $\hat{L}$ containing the left eigenvectors of the Lindbladian as rows, is as follows
\begin{equation}
\hat{L}=\left(\begin{array}{cccc}
               1 & 1 & 1 & 1 \\
               0 & -1 & 1 & 0 \\
               -\frac{2-f_+^{(1)}}{f_+^{(0)}} & -\frac{2-f_+^{(0)}-f_+^{(1)}}{2f_+^{(0)}} & -\frac{2-f_+^{(0)}-f_+^{(1)}}{2f_+^{(0)}} & 1 \\
               \frac{(2-f_+^{(0)})(2-f_+^{(1)})}{f_+^{(0)}f_+^{(1)}} & -\frac{2-f_+^{(0)}}{f_+^{(0)}} & -\frac{2-f_+^{(0)}}{f_+^{(0)}} & 1 \\
              \end{array}
\right)
\label{eq:and13} 
\end{equation}
The elements of these matrices $\hat{L}$ and $\hat{R}$ alongside the eigenvalues in Eq.~(\ref{eq:and9}) are used to calculate the explicit expressions of the time evolved density matrix elements in Eq.~(5) of the main text.

\section{Analysis of QMPE in density matrix elements using $\eta_{\mathrm{aq}}$ and $\eta_{\mathrm{bq}}$}
\label{phase-diagram}
In the main text, we stated the necessary criterion for the $\alpha$-th density matrix element to show QMPE is given by Eq.~(9).  
Using the expressions of right and left eigenvectors from Eqs.~(\ref{eq:and12})-(\ref{eq:and13}), we obtain a concise form of $S_\alpha$ given below
\begin{equation}
S_\alpha=g_\alpha(\eta_{\mathrm{aq}})\frac{\eta_{\mathrm{bq}}+1}{(1-\eta_{\mathrm{aq}})\eta_{\mathrm{bq}}+2}, \hspace*{0.2 cm}\alpha=1,2,3,4
\label{eq:ssalpha} 
\end{equation}
with $g_1(\eta_{\mathrm{aq}})=-2$, $g_2(\eta_{\mathrm{aq}})=g_3(\eta_{\mathrm{aq}})=\eta_{\mathrm{aq}}-1$ and $g_4(\eta_{\mathrm{aq}})=2\eta_{\mathrm{aq}}$. The important point to note in Eq.~(\ref{eq:ssalpha}) is the introduction of the two variables $\eta_{\mathrm{aq}}$ and $\eta_{\mathrm{bq}}$. Interestingly, $\eta_{\mathrm{aq}}$ is a combination of parameters {\it after quench} (aq) only, whereas $\eta_{\mathrm{bq}}$ is a function of parameters {\it before quench} (bq) only. The explicit expressions of $\eta_{\mathrm{aq}}$ and $\eta_{\mathrm{bq}}$ are given below in terms of the variables $f_{.}^{(.)}$ of the transition matrix $\hat{K}$ as
\begin{eqnarray}
\eta_{\mathrm{aq}}&=&\frac{2-f_{+}^{(0)}}{f_{+}^{(1)}},\cr
\eta_{\mathrm{bq}}&=&\frac{f_{+}^{(0),\mathrm{I}}f_{+}^{(1),\mathrm{I}}\left(4+2(f_{+}^{(0),\mathrm{II}}-f_{+}^{(1),\mathrm{II}})\right)-f_{+}^{(0),\mathrm{II}}f_{+}^{(1),\mathrm{II}}(4+2(f_{+}^{(0),\mathrm{I}}-f_{+}^{(1),\mathrm{I}}))}{f_{+}^{(0),\mathrm{I}}(2-f_{+}^{(1),\mathrm{I}})(4+2(f_{+}^{(0),\mathrm{II}}-f_{+}^{(1),\mathrm{II}}))-f_{+}^{(0),\mathrm{II}}(2-f_{+}^{(1),\mathrm{II}})\left(4+2(f_{+}^{(0),\mathrm{I}}-f_{+}^{(1),\mathrm{I}})\right)},
\label{eq:setagamma} 
\end{eqnarray}
where the superscripts $\mathrm{I}$ and $\mathrm{II}$ correspond to two different initial conditions discussed in the main text. Note that when one employs the criterion in Eq.~(9) [main text] to explore QMPE, in principle the parameter space under consideration is six dimensional, either in terms of the physically controllable parameters $(\mu_{\mathrm{L}}^{\mathrm{I}},\mu_{\mathrm{R}}^{\mathrm{I}},\mu^{\mathrm{II}},T_{\mathrm{i}};\mu,T)$ or equivalently in terms of the transition rates $(f_{+}^{(0),\mathrm{I}},f_{+}^{(1),\mathrm{I}},f_{+}^{(0),\mathrm{II}},f_{+}^{(1),\mathrm{II}};f_{+}^{(0)},f_{+}^{(1)})$ where these sets are connected by Eq.~(4) [main text]. However, Eqs.~(\ref{eq:ssalpha})-(\ref{eq:setagamma}) imply that instead of concentrating on the six dimensional parameter spaces, it would be sufficient to focus on the two-dimensional plane of $(\eta_{\mathrm{aq}},\eta_{\mathrm{bq}})$ that captures all the possibilities to get QMPE in $\rho_\alpha$-s.

Let us now proceed to classify different regions of the $\eta_{\mathrm{aq}}-\eta_{\mathrm{bq}}$ plane with distinct values of $\nu(\hat{\rho})$ (defined in the main text) that can take one of the four possible values $0,1,2,3$. From Eq.~(\ref{eq:setagamma}), we see that $\eta_{\mathrm{aq}}$ can only be positive since $f_{+}^{(0)}<2$ whereas $\eta_{\mathrm{bq}}$ can take both positive and negative values.

\subsection*{\underline{Case-1: $\eta_{\mathrm{aq}}<1$ and $\eta_{\mathrm{bq}}>0$}}
This case guarantees that $(\eta_{\mathrm{bq}}+1)/((1-\eta_{\mathrm{aq}})\eta_{\mathrm{bq}}+2)>0$. We must have $g_\alpha(\eta_{\mathrm{aq}})<0$ to satisfy the first criterion of Eq.~(9) [main text]. Clearly, $g_4(\eta_{\mathrm{aq}})$ and consequently $S_4$ does not satisfy this condition. The other three $g_1(\eta_{\mathrm{aq}}),g_2(\eta_{\mathrm{aq}})$ and $g_3(\eta_{\mathrm{aq}})$ satisfy it. Thus, the next step is to check if the corresponding magnitudes of $S_1,S_2,S_3
$ are less than $1$ or not. 

If we proceed to find out the condition under which $|S_1|<1$, that leads us to the following
\begin{equation}
\eta_{\mathrm{bq}}(1+\eta_{\mathrm{aq}})<0,
\label{eq:s1} 
\end{equation}
which is impossible for this case. Thus $S_1$ cannot exhibit QMPE under this case.

The condition $|S_2|<1$ gives rise to the condition
\begin{equation}
1+\eta_{\mathrm{aq}}>0,
\label{eq:s2} 
\end{equation}
which is valid under this case. We do not have to perform the similar procedure for $S_3$ since $S_3=S_2$.

Therefore, for Case-1, the singly occupied upspin state ($\uparrow$) and singly occupied downspin state ($\downarrow$) show QMPE and this parameter regime in the $\eta_{\mathrm{aq}}-\eta_{\mathrm{bq}}$ plane is characterized by $\nu(\hat{\rho})=2$.

\subsection*{\underline{Case-2: $\eta_{\mathrm{aq}}>1$ and $\eta_{\mathrm{bq}}>0$}}
In this case, the denominator $((1-\eta_{\mathrm{aq}})\eta_{\mathrm{bq}}+2)$ in Eq.~(\ref{eq:ssalpha}) can be either positive or negative. First we consider the sub-case where the denominator is positive i.e. $\eta_{\mathrm{bq}}<2/(\eta_{\mathrm{aq}}-1)$. In that case the term $(\eta_{\mathrm{bq}}+1)/\left((1-\eta_{\mathrm{aq}})\eta_{\mathrm{bq}}+2\right)>0$ which means $g_\alpha(\eta_{\mathrm{aq}})$ must be negative. This in turn means, $g_2(\eta_{\mathrm{aq}}),g_3(\eta_{\mathrm{aq}})$ and $g_4(\eta_{\mathrm{aq}})$ do not satisfy this condition and consequently $S_2,S_3$ and $S_4$ cannot exhibit QMPE under this case. Only chance remains for $S_1$. However, when we test if $|S_1|<1$ that gives
\begin{equation}
 \eta_{\mathrm{bq}}(1+\eta_{\mathrm{aq}})<0,  \label{eq:s3}                                                                                                                                                                                                                                                                                                                                                                   \end{equation}
which is impossible under this case. For $0<\eta_{\mathrm{bq}}<2/(\eta_{\mathrm{aq}}-1)$ there is no density matrix element showing QMPE. 

Let us consider the sub-case $\eta_{\mathrm{bq}}>2/(\eta_{\mathrm{aq}}-1)$. In this scenario, $g_\alpha(\eta_{\mathrm{aq}})$ must be positive. Thus, $S_1$ cannot show QMPE. The condition $|S_2|<1$ leads us to
\begin{equation}
1+\eta_{\mathrm{aq}}<0,
\label{eq:s4} 
\end{equation}
which is again impossible for the case under consideration. For $|S_4|<1$, one has to obey 
\begin{equation}
(\eta_{\mathrm{bq}}+2)(1+\eta_{\mathrm{aq}})<0,
\label{eq:s5} 
\end{equation}
which cannot be satisfied under this case.

Thus the parameter regime for Case-2, does not exhibit QMPE for any of the density matrix elements and it is characterized by $\nu(\hat{\rho})=0$.

\subsection*{\underline{Case-3: $\eta_{\mathrm{aq}}<1$ and $\eta_{\mathrm{bq}}<0$}}
For this case, we rewrite the definition of $S_\alpha$ in Eq.~(\ref{eq:ssalpha}) as
\begin{equation}
S_\alpha=g_\alpha(\eta_{\mathrm{aq}})\frac{1-|\eta_{\mathrm{bq}}|}{2-(1-\eta_{\mathrm{aq}})|\eta_{\mathrm{bq}}|}.
\label{eq:s61} 
\end{equation}
Let us first consider the sub-case when $|\eta_{\mathrm{bq}}|<\mathrm{min}[1,2/(1-\eta_{\mathrm{aq}})]$ i.e $|\eta_{\mathrm{bq}}|<1$, where $\mathrm{min}[a,b]$ selects the smaller one between $a$ and $b$. In this sub-case, $g_\alpha(\eta_{\mathrm{aq}})$ must be negative. This implies $S_4$ cannot show QMPE. If we consider $|S_1|<1$, that leads us to
\begin{equation}
(1+\eta_{\mathrm{aq}})|\eta_{\mathrm{bq}}|>0,
\label{eq:s6} 
\end{equation}
which is true. So, $S_1$ shows QMPE. For $|S_2|<1,$ we have to satisfy
\begin{equation}
1+\eta_{\mathrm{aq}}>0
\label{eq:s7}
\end{equation}
which is valid. As a consequence, both $S_2$ and $S_3$ show QMPE.

Therefore, for the sub-case $\eta_{\mathrm{aq}}<1$ and $\eta_{\mathrm{bq}}<0$ and $|\eta_{\mathrm{bq}}|<1$, the doubly occupied state ($d$), the singly occupied upspin ($\uparrow$) and singly occupied downspin ($\downarrow$) state exhibit QMPE and this parameter regime is characterized by $\nu(\hat{\rho})=3$.\\

Next we consider the sub-case where $1<|\eta_{\mathrm{bq}}|<2/(1-\eta_{\mathrm{aq}})$. In this sub-case the numerator of the fraction in Eq.~(\ref{eq:s61}) is negative while its denominator is positive. So, $g_\alpha(\eta_{\mathrm{aq}})$ has to be positive. This means $S_1,S_2$ and $S_3$ cannot show QMPE under this sub-case. For $|S_4|<1$, we arrive at the condition below
\begin{equation}
|\eta_{\mathrm{bq}}|<2.
\label{eq:s8} 
\end{equation}
For $\eta_{\mathrm{aq}}<1$ and $\eta_{\mathrm{bq}}<0$ and $1<|\eta_{\mathrm{bq}}|<2$, only the empty  state ($e$) exhibits QMPE and this region is characterized by $\nu(\hat{\rho})=1$.\\

From the analysis of the previous sub-case, it is also evident that for $\eta_{\mathrm{aq}}<1$ and $\eta_{\mathrm{bq}}<0$ and $2<|\eta_{\mathrm{bq}}|<2/(1-\eta_{\mathrm{aq}})$, no density matrix element exhibit QMPE and we have $\nu(\hat{\rho})=0$. \\

Finally, we consider the sub-case $|\eta_{\mathrm{bq}}|>2/(1-\eta_{\mathrm{aq}})$. Here the Eq.~(\ref{eq:ssalpha}) can be rewritten as
\begin{equation}
S_\alpha=g_\alpha(\eta_{\mathrm{aq}})\frac{|\eta_{\mathrm{bq}}|-1}{(1-\eta_{\mathrm{aq}})|\eta_{\mathrm{bq}}|-2}.
\label{eq:s9} 
\end{equation}
Thus $g_\alpha(\eta_{\mathrm{aq}})$ has to be negative. This means $S_4$ cannot show QMPE. To have $|S_1|<1$, one has to satisfy the following condition
\begin{equation}
(1+\eta_{\mathrm{aq}})|\eta_{\mathrm{bq}}|<0,
\label{eq:s10} 
\end{equation}
which is not possible. The criterion $|S_2|<1$ leads us to
\begin{equation}
1+\eta_{\mathrm{aq}}<0,
 \label{eq:s11}
\end{equation}
which cannot be satisfied. So, for $|\eta_{\mathrm{bq}}|>2/(1-\eta_{\mathrm{aq}})$, no density matrix element shows QMPE.
\begin{figure}[h]
\centering \includegraphics[width=8.6 cm]{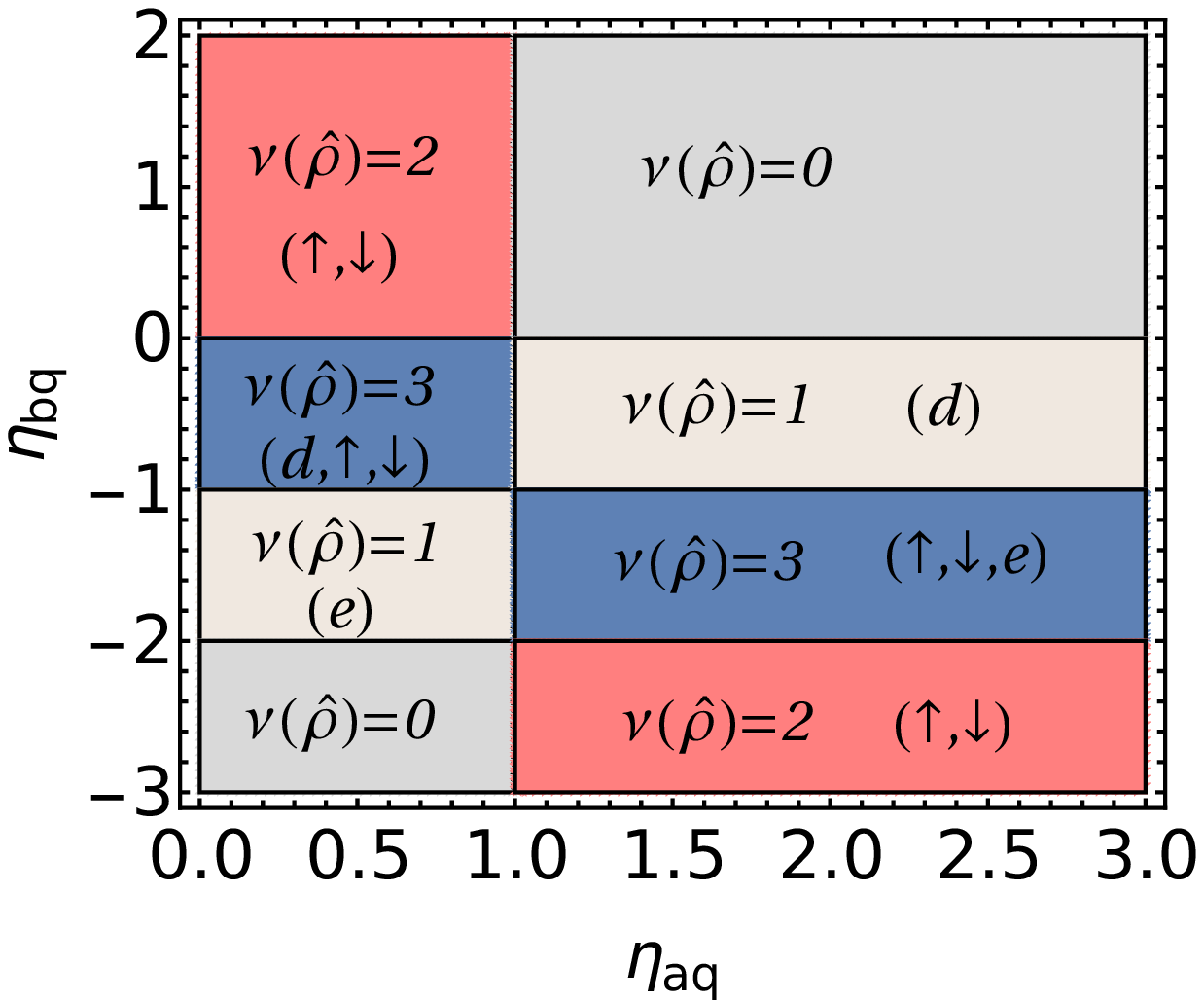}
\caption{The figure illustrates the possibility of QMPE in density matrix elements ($d$ and $e$ correspond to doubly occupied and empty states, respectively) characterized by $\nu(\hat{\rho})$ in the $\eta_{\mathrm{aq}}-\eta_{\mathrm{bq}}$ plane. We observe four distinct regimes  and importantly, we get the information regarding which region in the parameter space would give rise to what value of $\nu(\hat{\rho})$..}
\label{fig:eta}
\end{figure}

The last two sub-cases together imply that, the parameter regime $\eta_{\mathrm{aq}}<1$ and $\eta_{\mathrm{bq}}<0$ and $|\eta_{\mathrm{bq}}|>2$ is characterized by $\nu(\hat{\rho})=0$.

\subsection*{\underline{Case-4: $\eta_{\mathrm{aq}}>1$ and $\eta_{\mathrm{bq}}<0$}}
For this case, we would like to express Eq.~(\ref{eq:ssalpha}) as
\begin{equation}
S_\alpha=g_\alpha(\eta_{\mathrm{aq}})\frac{1-|\eta_{\mathrm{bq}}|}{(\eta_{\mathrm{aq}}-1)|\eta_{\mathrm{bq}}|+2}.
\label{eq:s12} 
\end{equation}
We first consider the sub-case $|\eta_{\mathrm{bq}}|<1$. Under this sub-case, we must have $g_\alpha(\eta_{\mathrm{aq}})<0$ for QMPE to happen. As a consequence  $S_2,S_3$ and $S_4$ cannot exhibit QMPE. The criterion $|S_1|<1$ leads us to
\begin{equation}
(1+\eta_{\mathrm{aq}})|\eta_{\mathrm{bq}}|>0,
\label{eq:s13} 
\end{equation}
which is true. So, for $\eta_{\mathrm{aq}}>1$, $\eta_{\mathrm{bq}}<0$ and $|\eta_{\mathrm{bq}}|<1$, only the doubly occupied state shows QMPE and this parameter regime is characterized by $\nu(\hat{\rho})=1$.\\

Next we consider the sub-case where $|\eta_{\mathrm{bq}}|>1$. For this sub-case, we must have $g_\alpha(\eta_{\mathrm{aq}})>0$ for QMPE to occur. This implies that $S_1$ cannot show QMPE under this sub-case. For $|S_2|<1$, one has to satisfy
\begin{equation}
1+\eta_{\mathrm{aq}}>0,
\label{eq:s14} 
\end{equation}
which is true. So, upspin and downspin states show QMPE. The remaining criterion $|S_4|<1$ gives rise to the condition below
\begin{equation}
|\eta_{\mathrm{bq}}|<2.
\label{eq:s15} 
\end{equation}
The above analysis implies that, for $\eta_{\mathrm{aq}}>1$, $\eta_{\mathrm{bq}}<0$ and $1<|\eta_{\mathrm{bq}}|<2$, three density matrix elements corresponding to the states $\uparrow,\downarrow,e$ exhibit QMPE and the parameter regime is characterized by $\nu(\hat{\rho})=3$. Note that we also obtained $\nu(\hat{\rho})=3$ for a different sub-case under Case-3 where $\eta_{\mathrm{aq}}<1$, $\eta_{\mathrm{bq}}<0$ and $|\eta_{\mathrm{bq}}|<1$, however the doubly occupied state in that case shows QMPE whereas it is the empty state that shows QMPE for the present sub-case.

We also understand that for $\eta_{\mathrm{aq}}>1$, $\eta_{\mathrm{bq}}<0$ and $|\eta_{\mathrm{bq}}|>2$, only the upspin and downspin states show QMPE, characterizing this parameter regime by $\nu(\hat{\rho})=2$. \\

Thus, we have studied in details how to identify different regimes in the $\eta_{\mathrm{aq}}-\eta_{\mathrm{bq}}$ plane with distinct values of $\nu(\hat{\rho})$. We have summarized the whole classification of QMPE in $\eta_{\mathrm{aq}}-\eta_{\mathrm{bq}}$ plane in Fig.~\ref{fig:eta}. We should mention, although the figure describes the possibility of QMPE up to finite values of $\eta_{\mathrm{aq}}$ and $\eta_{\mathrm{bq}}$, these regions can be simply extended to $\eta_{\mathrm{aq}}<\infty$ and $-\infty<\eta_{\mathrm{bq}}<\infty$ without any further calculation. Also, note that the parameter regions exhibiting $\nu(\hat{\rho})=3,1$ are much narrower in comparison to the parameter regions showing $\nu(\hat{\rho})=2,0$. In fact, in an infinite plane sheet of $\eta_{\mathrm{aq}}-\eta_{\mathrm{bq}}$, the parameter region exhibiting $\nu(\hat{\rho})=3,1$ appears as an infinitely long band of extremely narrow width. This is the reason why we observed such narrow regions showing $\nu(\hat{\rho})=3,1$ in Fig.~1 of the main text as a function of the physical control parameters. 

\section{Mixed thermal QMPE and inverse  thermal QMPE}
\label{mixed-inverse}
\begin{figure}[h]
  \centering
  \subfigure[]{\includegraphics[scale=0.45]{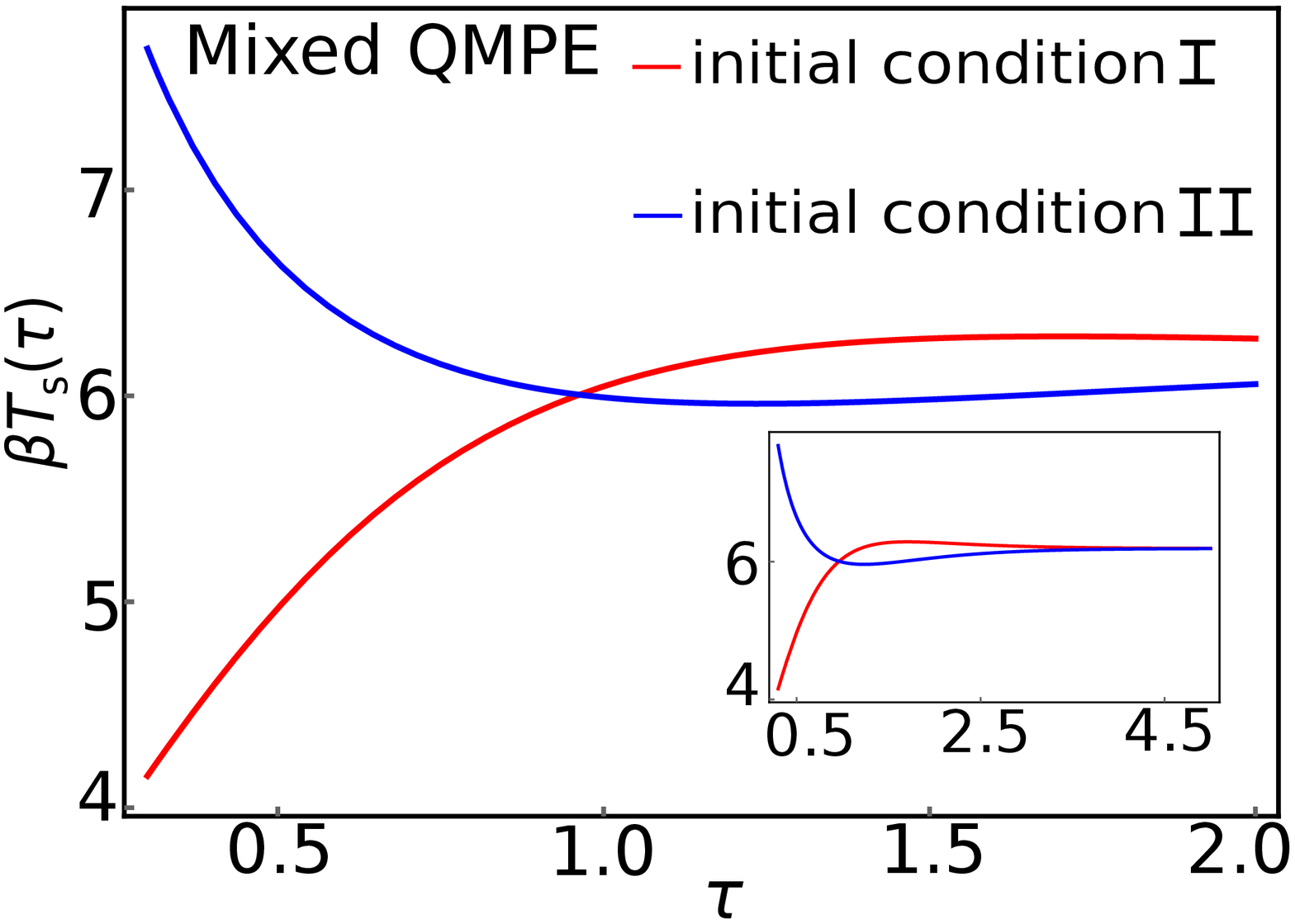}}\hfill
  \subfigure[]{\includegraphics[scale=0.45]{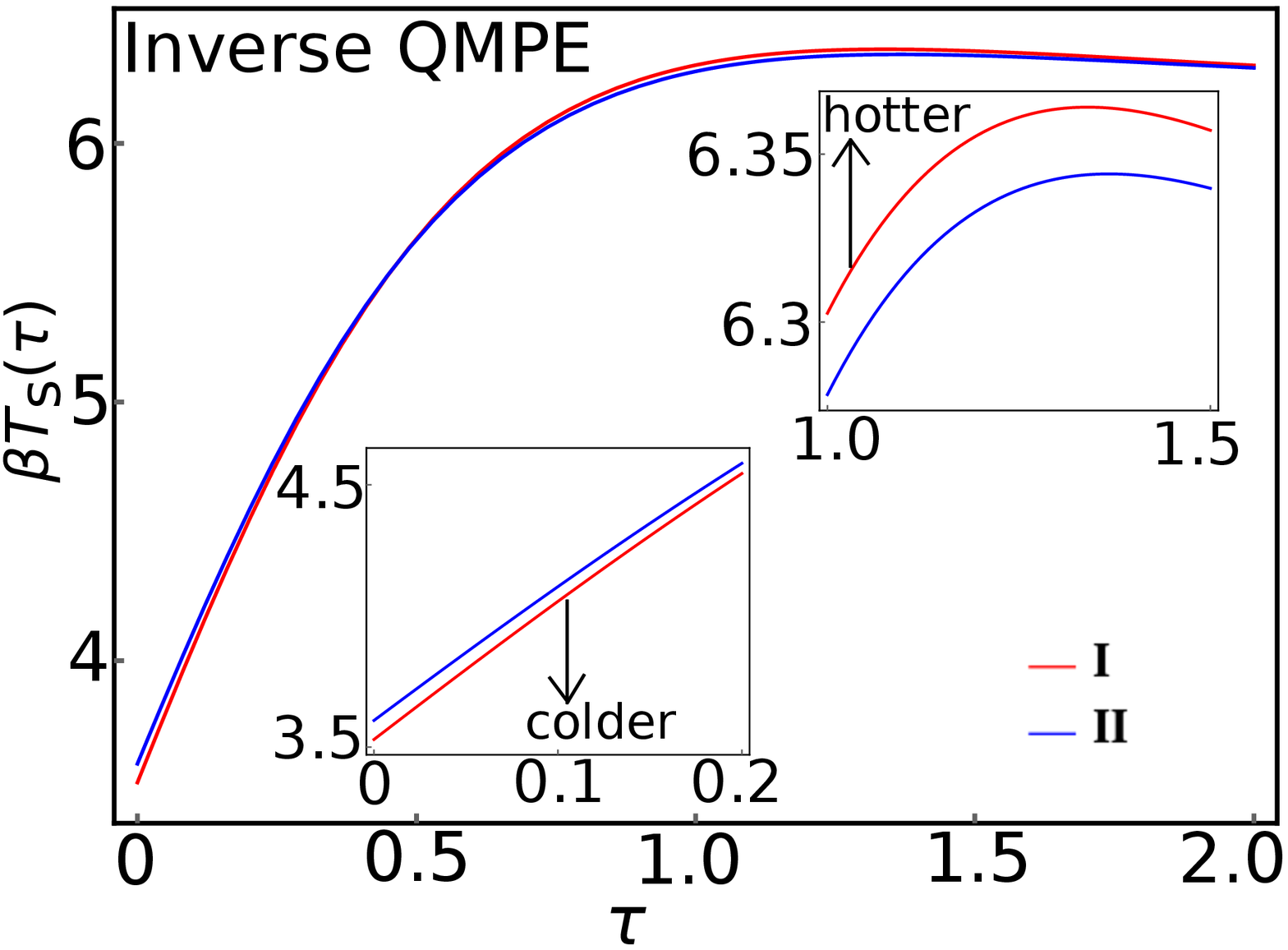}}
  \caption{The figure (a) illustrates the mixed thermal QMPE where one initial temperature is lower and the other is higher  than the steady value. The inset confirms the convergence of both temperatures to same steady state. Parameters used are $\beta\epsilon_0=2.0, \beta U=1.25, \beta\mu_{\mathrm{L}}^{\mathrm{I}}=1.5, \beta\mu_{\mathrm{R}}^{\mathrm{I}}=1.0, \beta\mu^{\mathrm{II}}=2.43, \beta T_\mathrm{i}=1.15, \beta\mu=2.0$. The figure (b) shows the inverse QMPE where both initial temperatures are lower than the steady value. Since the  temperature trajectories are very close in this case, we use the two insets for clear visibility of the role reversals (i.e. hotter becoming colder and vice versa) to generate QMPE. Parameters used are $\beta\epsilon_0=2.0, \beta U=1.25, \beta\mu_{\mathrm{L}}^{\mathrm{I}}=1.9, \beta\mu_{\mathrm{R}}^{\mathrm{I}}=1.0, \beta\mu^{\mathrm{II}}=1.5, \beta T_\mathrm{i}=1.15, \beta\mu=2.0$.}
  \label{fig:mixedinv}
\end{figure}
In the main text, we have explicitly shown the {\it normal} thermal QMPE in the QD where {\it both initial temperatures are higher than the steady state value} and during their cooling relaxation, they cross each other and interchange their identities (i.e. hotter becomes colder and vice versa) to produce QMPE. In this appendix, we present examples of two other types of thermal QMPE occurring in the QD, namely {\it mixed} thermal QMPE [Fig.~\ref{fig:mixedinv}(a)] and {\it inverse} thermal QMPE [Fig.~\ref{fig:mixedinv}(b)].

In Fig.~\ref{fig:mixedinv}(a), we observe that one of the initial temperatures (I) is lower while the other initial temperature (II) is higher than the steady state value. As a consequence, one trajectory cools down and the other one heats up towards the steady value. Therefore the QMPE generated (by the crossing of the temperatures' trajectories) in this case can be recognized as the {\it mixed} thermal QMPE as per the nomenclature in existing literature. The inset shows the convergence of the two different initial conditions to the same steady value confirming thermalization in the system. On the other hand, in Fig.~\ref{fig:mixedinv}(b), both initial temperatures are lower than their steady state value. Both of them heat up towards the steady value and cross each other during the relaxation process to create QMPE. Since the scenario is opposite to the normal QMPE, this case falls in the category of {\it inverse} thermal QMPE. Based on the examples discussed in the main text (Fig.~3) and the supplemental material (Fig.~\ref{fig:mixedinv}(a) and Fig.~\ref{fig:mixedinv}(b)), we see that the inverse QMPE [Fig.~\ref{fig:mixedinv}(b)] is weaker in comparison to normal QMPE (Fig.~3 [main text]) and mixed QMPE [Fig.~\ref{fig:mixedinv}(a)], in the sense that the  temperatures for the two initial conditions remain very close to each other for inverse QMPE.

\section{Retrieving classical limit of the system temperature}
\label{classical}
The validity of the system temperature (Eq.~(11) [main text]) used to demonstrate QMPE in the QD, should be carefully checked. One way to investigate this is to see if the temperatures starting from different initial conditions converge to the same steady state temperature and thereby assure the thermalization of the system. Indeed, we have shown explicitly in the inset of Fig.~3 [main text], inset of Fig.~\ref{fig:mixedinv}(a) and Fig.~\ref{fig:mixedinv}(b) that  temperatures starting from distinct initial conditions lead to same steady state value. In this appendix, we examine the reliability of the temperature in another way by checking the classical limit of the temperature.
\begin{figure}[h]
  \centering
  \subfigure[]{\includegraphics[scale=0.475]{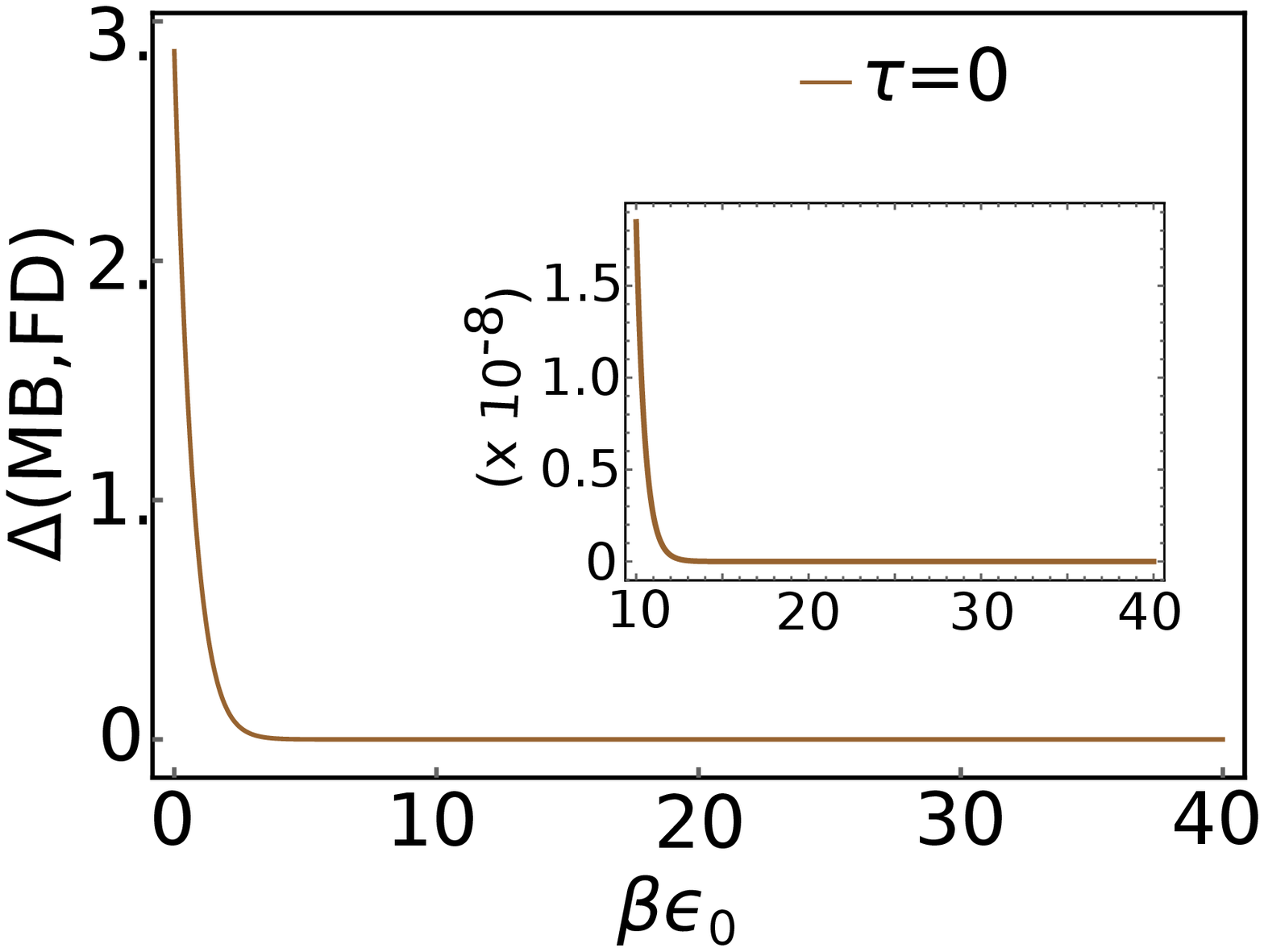}}\hfill
  \subfigure[]{\includegraphics[scale=0.5]{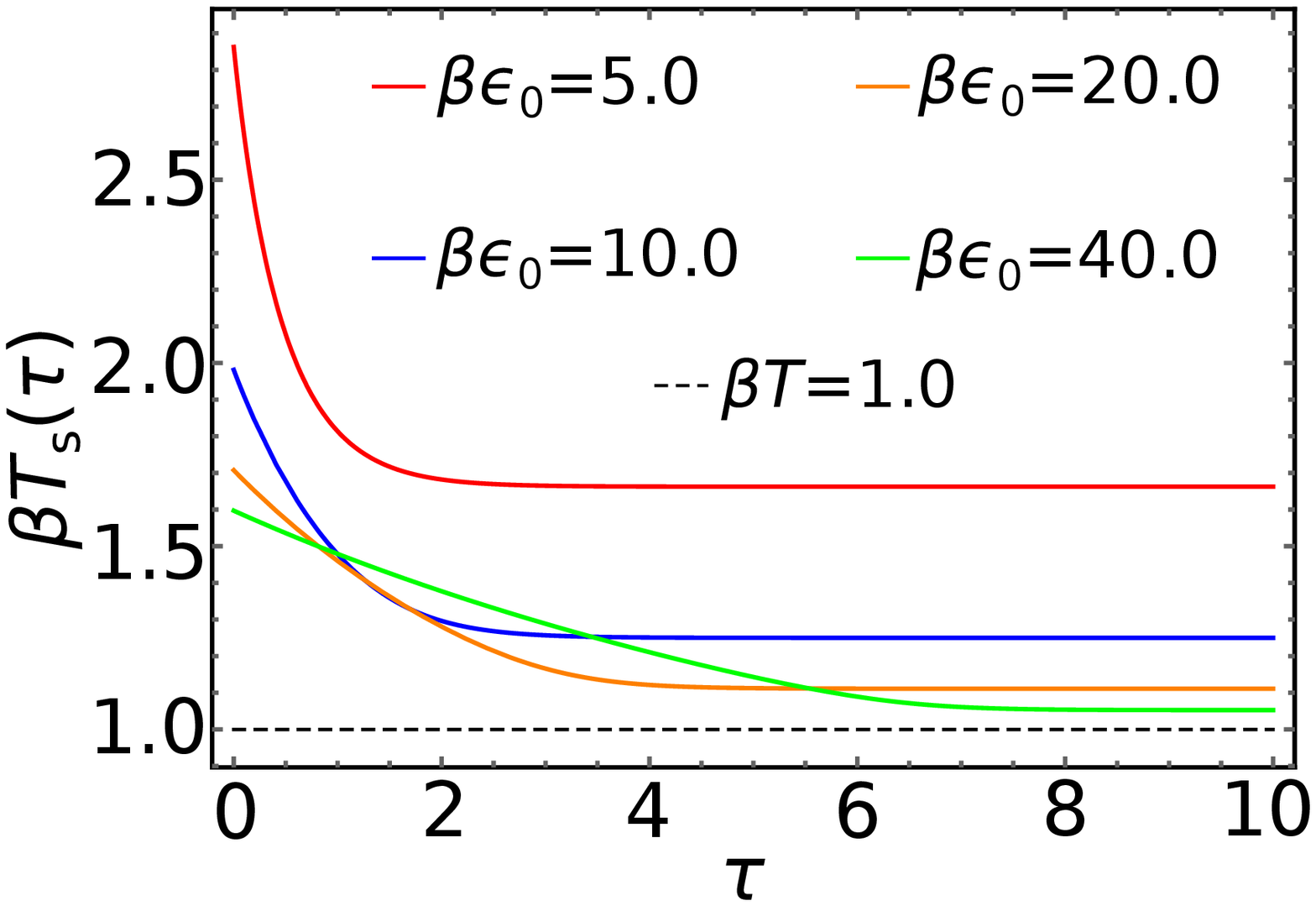}}
  \caption{The figure (a) shows how the classical limit can be set up at $\tau=0$ by increasing $\beta\epsilon_0$. The figure (b) shows that the steady state ($\tau\rightarrow\infty$) value of temperature converges to the initial reservoir temperature (dotted line) with increasing $\beta\epsilon_0$. Such equilibration between system-reservoir indicates the recovery of classical behavior at large $\beta\epsilon_0$ and points towards the correctness of the definition of system temperature. Parameters used are $\beta U=1.25, \beta\mu_{\mathrm{L}}^{\mathrm{I}}=4.5, \beta\mu_{\mathrm{R}}^{\mathrm{I}}=1.0, \beta\mu^{\mathrm{II}}=2.43, \beta T_\mathrm{i}=5.0, \beta\mu=2.0$.}
  \label{fig:classical}
\end{figure}
To do so, we first revisit the definitions of the parameters $f_+^{(j)}$ with $j=0,1$ from Eq.~(4) [main text]. These basically consist of the Fermi-Dirac (FD) distributions of the form $\frac{1}{1+e^{X}}$ where $X:=\beta(\epsilon_0+jU-\mu)$ where $j=0,1$ and $\beta=1/T$. Note that one can retrieve the classical Maxwell-Boltzmann (MD) distribution of the form $e^{-X}$ by making $e^{X}\gg1$. In our case, one possible way of getting close to the classical MB distribution is to increase $\beta\epsilon_0$ so that the difference $\beta(\epsilon_0-\mu)$ increases. We would like to check this prescription by defining the following measure of difference between FD and MB distributions,
\begin{equation}
\Delta(\mathrm{MB},\mathrm{FD})=2e^{-X}-2/(1+e^{X}),
\label{eq:delta} 
\end{equation}
where the factor $2$ in $f_+^{(j)}$ comes from the simple choice $\mu_{\mathrm{L}}=\mu_{\mathrm{R}}$ (like initial condition II used in the main text, similar analysis can be straightforwardly extended to $\mu_{\mathrm{L}}\neq\mu_{\mathrm{R}}$ case). In Fig.~\ref{fig:classical}(a), we observe that $\Delta(\mathrm{MB},\mathrm{FD})$ goes to zero monotonically as we increase $\beta\epsilon_0$ and for considerably large values of $\beta\epsilon_0$ we recover the classical behavior of the reservoirs and the system initially at $\tau=0$. To understand what it means for the system temperature $\beta T_\mathrm{s}(\tau)$, we present the temporal variations of $\beta T_\mathrm{s}(\tau)$ in Fig.~\ref{fig:classical}(b) for different values of $\beta\epsilon_0$. Indeed we observe that the steady state system temperature (at $\tau\rightarrow\infty$) gets more and more closer to the initial ($\tau=0$) reservoir temperature $\beta T$ (i.e. unity) as we increase $\beta\epsilon_0$ i.e. move towards the classical equilibration between system and reservoir. Thus, the temperature we have used for the quantum mechanical system, indeed recovers the classically expected steady state behavior in the classical limit when the FD distributions can be approximated as MB distributions. This assures the validity of the system temperature we utilize to study thermal QMPE.

\section{QMPE in energy, von Neumann entropy and Kullback-Leibler divergence}
\label{energy-entropy}
In this appendix we discuss that apart from the system  temperature, there are other observables like average energy, von-Neumann entropy, Kullback-Leibler divergence  of the system that also exhibit QMPE. The average energy $E_\mathrm{s}(\tau)$ and the  von-Neumann entropy $S_{\mathrm{vN}}(\tau)$ are calculated as
\begin{eqnarray}
E_\mathrm{s}(\tau)&=&\mathrm{Tr}[\hat{\rho}(\tau)\hat{H}_\mathrm{s}],\cr
S_\mathrm{vN}(\tau)&=& -\sum_{\alpha=1}^{4} \rho_\alpha(\tau)\mathrm{ln}(\rho_\alpha(\tau)).
\label{eq:enent} 
\end{eqnarray}
\begin{figure}[h]
  \centering
  \subfigure[]{\includegraphics[scale=0.475]{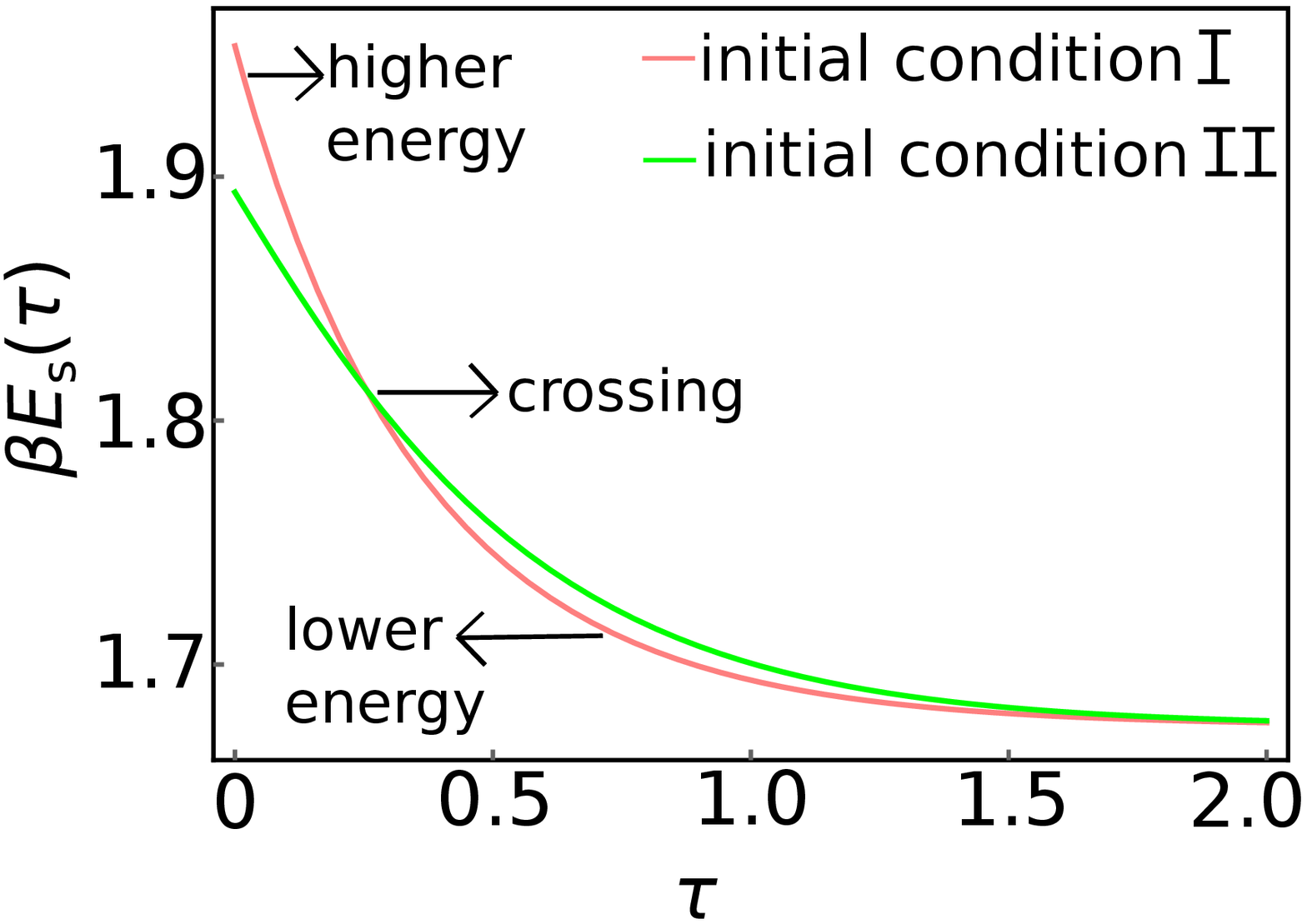}}\hfill
  \subfigure[]{\includegraphics[scale=0.475]{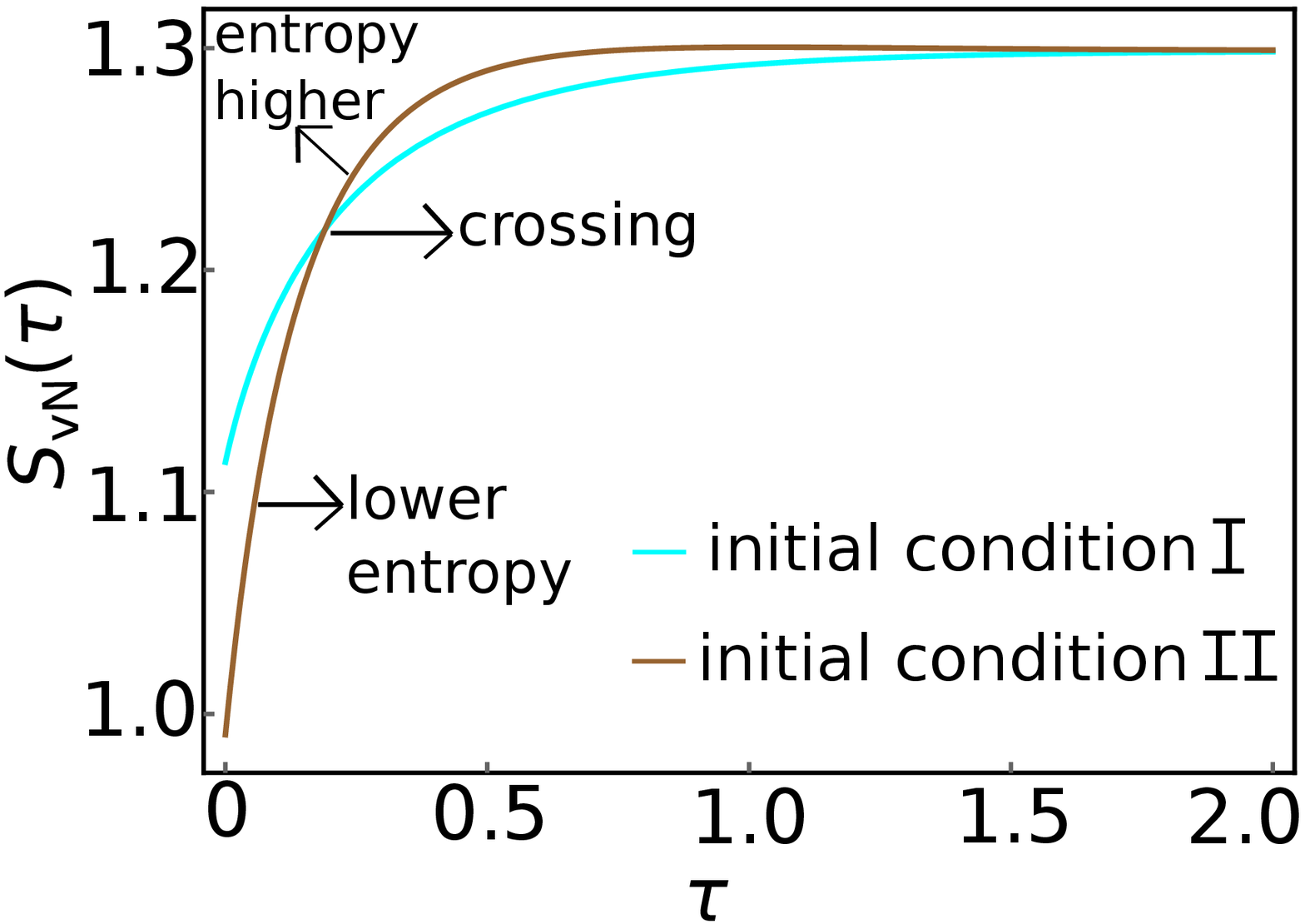}}
\caption{The figure (a) illustrates normal QMPE in energy (scaled by $\beta$) where both initial conditions have higher energies than their steady state value and cross each other at some intermediate finite time before reaching steady state. Parameters used are $\beta\epsilon_0=2.0, \beta U=1.25, \beta\mu_{\mathrm{L}}^{\mathrm{I}}=3.47, \beta\mu_{\mathrm{R}}^{\mathrm{I}}=1.0, \beta\mu^{\mathrm{II}}=2.43, \beta T_\mathrm{i}=0.25, \beta\mu=2.0$. The figure (b) exhibits inverse QMPE in von-Neumann entropy where both initial conditions have lower entropy than their steady state value and cross each other at finite time. Parameters used are $\beta\epsilon_0=2.0, \beta U=1.25, \beta\mu_{\mathrm{L}}^{\mathrm{I}}=2.4, \beta\mu_{\mathrm{R}}^{\mathrm{I}}=1.0, \beta\mu^{\mathrm{II}}=2.43, \beta T_\mathrm{i}=0.25, \beta\mu=2.0$.}
  \label{fig:enent}
\end{figure}
In Fig.~\ref{fig:enent}(a), we provide an example of QMPE in  energy (scaled by $\beta$) of the system. We observe that starting from two different initial conditions, the initially higher energy trajectory (mimicing initial hotter system) loses energy faster and crosses the  initially lower energy trajectory (mimicing initial colder system) at some finite time and thereafter reverse their identities (system with higher initial energy becomes lower energy system and vice versa). In connection to the classifications of QMPE as normal, inverse and mixed; this example of QMPE in $E_\mathrm{s}(\tau)$ can be identified as normal QMPE since both the initial conditions have higher energies than their steady state value. In Fig.~\ref{fig:enent}(b), we present QMPE in von-Neumann entropy. Starting from two distinct initial conditions, the initially lower entropy trajectory gains entropy at a faster rate and crosses the initially higher entropy trajectory and consequently reverse their roles (system with lower initial entropy becomes higher entropy system and vice versa). This QMPE in $S_\mathrm{vN}(\tau)$ is inverse QMPE in nature because the initial entropies for both initial conditions are lower than their steady state value.

One of the most commonly used tool to analyze MPE or QMPE in the existing literature is the Kullback-Leibler (KL) divergence.  It is a dynamic measure of distance-from-steady state defined below,
\begin{equation}
D_{\mathrm{KL}}(\tau)=\mathrm{Tr}[\hat{\rho}(\tau)\,\left(\mathrm{ln}\hat{\rho}(\tau)-\mathrm{ln}\hat{\rho}_{\mathrm{ss}}\right)],
\label{eq:dkl} 
\end{equation}
where $\hat{\rho}_{\mathrm{ss}}$ corresponds to the steady state density matrix. Since the density matrix in our case does not contain any off diagonal elements we can rewrite the expression for $D_{\mathrm{KL}}(\tau)$ in terms of density matrix elements as
\begin{eqnarray}
D_{\mathrm{KL}}(\tau)=\sum_{\alpha=1}^{4}\rho_\alpha(\tau)\,\mathrm{ln}\left(\frac{\rho_\alpha(\tau)}{\rho_{\mathrm{ss},\alpha}}\right)=-\sum_{\alpha=1}^{4}\rho_\alpha(\tau)\,\mathrm{ln}(\rho_{\mathrm{ss},\alpha})-S_{\mathrm{vN}},
\label{eq:dkl1} 
\end{eqnarray}
where $S_{\mathrm{vN}}$ is the von-Neumann entropy defined in Eq.~(\ref{eq:enent}).
\begin{figure}[t]
  \centering
  \subfigure[]{\includegraphics[scale=0.537]{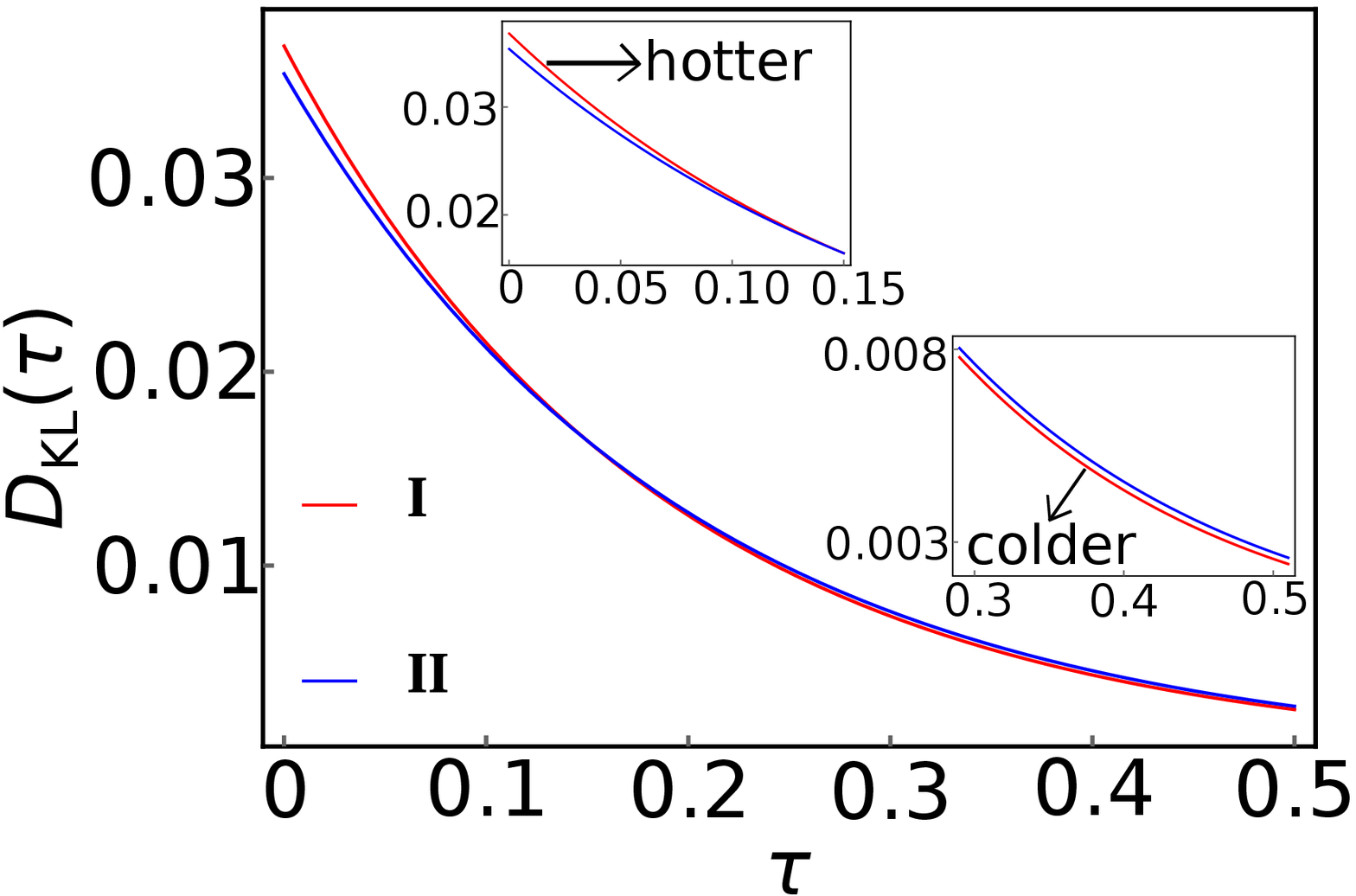}}\hfill
  \subfigure[]{\includegraphics[scale=0.463]{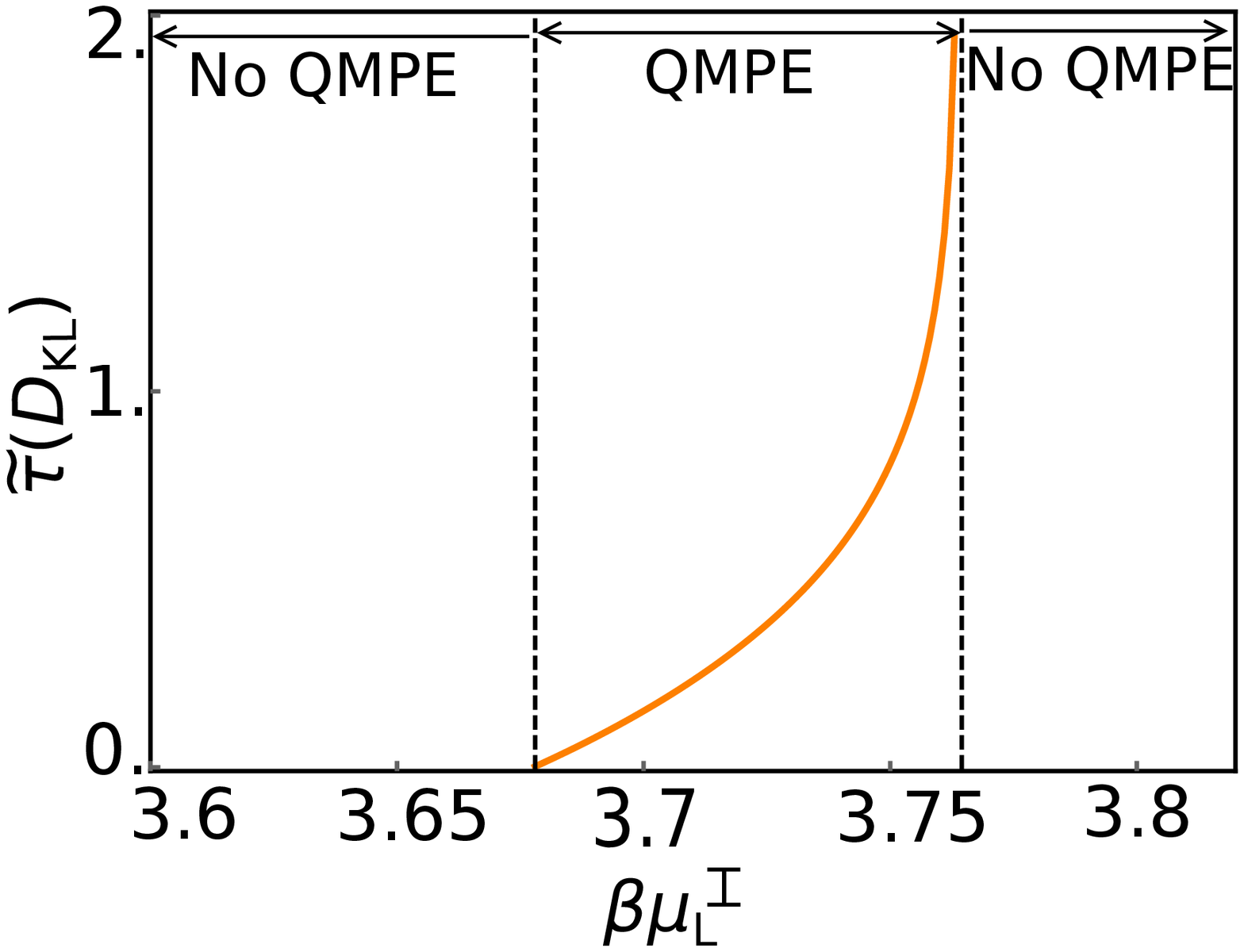}}
  \caption{The figure (a) illustrates QMPE in KL divergence where the trajectory with initially higher $D_{\mathrm{KL}}$ value crosses the other trajectory (with initially lower $D_{\mathrm{KL}}$ value) at some finite time and thereby reverses its identity by becoming the trajectory with lower $D_{\mathrm{KL}}$ value. Since the QMPE is weak in magnitude, we use insets for clear visibility. The figure (b) shows the variation of the temporal order parameter $\tilde{\tau}(D_{\mathrm{KL}})$ [Eq.~(\ref{eq:tau_dkl})] as a function of the control parameter $\beta\mu_{{\mathrm{L}}}^{\mathrm{I}}$. Only a narrow region in the parameter space shows QMPE in $D_{\mathrm{KL}}(\tau)$.  Common set of parameters used for both figures are $\beta\epsilon_0=2.0, \beta U=1.25, \beta\mu_{\mathrm{R}}^{\mathrm{I}}=1.0, \beta\mu^{\mathrm{II}}=2.43, \beta T_\mathrm{i}=1.15, \beta\mu=2.0$, additionally for (a) $\beta\mu_{\mathrm{L}}^{\mathrm{I}}=3.7$.}
  \label{fig:dkl}
\end{figure}

To investigate QMPE in $D_{\mathrm{KL}}(\tau)$, we consider as always, two different initial conditions I and II. In Fig.~\ref{fig:dkl}(a), we observe that the trajectory with initially higher value of $D_{\mathrm{KL}}(\tau)$ (thereby identified as initially hotter) crosses the other trajectory with initially lower value of $D_{\mathrm{KL}}(\tau)$ (thereby identified as initially colder) and thereafter reverse their identities (hotter becoming colder and vice versa) to exhibit QMPE. One can equivalently focus on the differences between these trajectories $\Delta D_{\mathrm{KL}}(\tau)$ expressed as
\begin{equation}
\Delta D_{\mathrm{KL}}(\tau)=D_{\mathrm{KL}}^{\mathrm{I}}(\tau)-D_{\mathrm{KL}}^{\mathrm{II}}(\tau)=-\sum_{\alpha=1}^{4}\Delta\rho_\alpha(\tau)\,\mathrm{ln}(\rho_{\mathrm{ss},\alpha})-\Delta S_{\mathrm{vN}},
\label{eq:ddkl} 
\end{equation}
such that $\Delta D_{\mathrm{KL}}(\tau)$ must become zero at some finite time and change sign before reaching steady state in order to produce QMPE. In analogy to QMPE in temperature and density matrix elements discussed in the main text, we would like to define the following temporal order parameter $\tilde{\tau}(D_{\mathrm{KL}})$ to characterize the presence and absence of QMPE in $D_{\mathrm{KL}}$ in the control parameter space,
\begin{eqnarray}
0<\tilde{\tau}(D_{\mathrm{KL}})<\infty&:& \hspace*{0.2 cm} \mathrm{QMPE}\,\,\mathrm{in}\,\,\mathrm{KL}\,\,\mathrm{divergence},\cr
\tilde{\tau}(D_{\mathrm{KL}})\rightarrow\infty&:& \hspace*{0.2 cm} \mathrm{no}\,\,\mathrm{QMPE},
\label{eq:tau_dkl} 
\end{eqnarray}
where $\tilde{\tau}(D_{\mathrm{KL}})$ satisfies $\Delta D_{\mathrm{KL}}(\tau)=0$. In Fig.~\ref{fig:dkl}(b), we present the variation of $\tilde{\tau}(D_{\mathrm{KL}})$ as a function of the tuning parameter $\beta\mu_{{\mathrm{L}}}^{\mathrm{I}}$. We observe that only a narrow range of $\beta\mu_{{\mathrm{L}}}^{\mathrm{I}}$ is able to result in QMPE. Note that we have purposefully used the exact same set of parameters to compare the QMPE in KL divergence [Fig.~\ref{fig:dkl}(b)] to that of thermal QMPE (Fig.~4 [main text]). Surprisingly, the parameter region leading to QMPE in KL divergence corresponds to no thermal QMPE. Similarly, the parameter ranges giving rise to thermal QMPE correspond to absence of QMPE in $D_{\mathrm{KL}}$. This implies that, there are parameter regions for which $D_{\mathrm{KL}}$ cannot be an alternative indicator for thermal QMPE. Deeper understanding of connections between the occurrence of QMPE in different entities require further works.

\end{document}